\documentclass[aps,preprint]{revtex4}%
\usepackage{amsfonts}
\usepackage{amsmath}
\usepackage{amssymb}
\usepackage{graphicx}
\usepackage{natbib}%
\usepackage{xr}
\usepackage[caption=false]{subfig}
\usepackage{textcomp}
\providecommand{\U}[1]{\protect\rule{.1in}{.1in}}
\DeclareMathOperator{\sinhInt}{sinhInt}
\DeclareMathOperator{\expIntEi}{expIntEi}
\DeclareMathOperator{\Det}{Det}
\begin{document}
\preprint{ }
\title{Strong coupling electrostatic theory of polymer counterions close to planar charges}

\author{Sandipan Dutta}
\affiliation{ Asia Pacific Center for Theoretical Physics, Pohang, Gyeongbuk, 790-784, Korea}
\author{and Y.S. Jho}
\email{ysjho@apctp.org}
\affiliation{Department of Physics, Pohang University of Science and Technology, Asia Pacific Center for Theoretical Physics, Pohang, Gyeongbuk, 790-784, Korea}

\begin{abstract}
Strong coupling phenomena, such as the like charged macroions attraction, opposite charged macroions repulsion, charge renormalization or charge inversion, are known to be mediated by 
multivalent counterions. Most theories treat the counterions as point charges, and describe the system by a single coupling parameter that measures 
the strength of the Coulomb interactions. In many biological systems, the counterions are highly charged and have finite sizes and can be well-described by polyelectrolytes.
The shapes and orientations of these polymer counterions play a major role in the thermodynamics of these systems. In this work we apply a field theoretic description in the strong 
coupling regime to polyelectrolytes. We work out the special cases of rod-like polymer counterions confined by one and two charged walls respectively.
The effects of the geometry of the rod-like counterions and the excluded volume of the walls on the density, pressure and the free energy of the 
rodlike counterions are discussed.
\end{abstract}

%\keywords{}

\maketitle

\section{Introduction}
Over the last couple of decades, several experiments have been performed on the peculiar phenomena of like-charged attractions in biological and soft materials, 
that causes the condensation of highly charged DNA ~\cite{matulis2002thermodynamics,radler1997structure, koltover1999phase, wong2000hierarchical,
pelta1996dna,raspaud1998precipitation}, the aggregation of viruses ~\cite{li2011superlattice,butler2003ion} and network formation in actin filaments
~\cite{angelini2003like}. These attractions are often mediated by multivalent ions like the aggregation of $M13$ viruses by divalent diamine ions
\cite{butler2003ion}, the condensation of DNA \cite{teif2011condensed, bloomfield1996dna} or the folding of RNA \cite{draper2005ions} by $Mg^{2+}$.

The counter-intuitive phenomena of like charge attractions is of great interest in theoretical field 
because they can not be explained within the framework of the mean-field Poisson-Boltzmann theory. 
Guldbrand et al. were the first to suggest a molecular mechanism for these attractions by the counterion correlations and 
fluctuations~\cite{guldbrand1984electrical,guldbrand1986monte}. 
This idea was further extended by Shklovskii based on strongly correlations among the counterions condensed into a Wigner crystal 
around a fixed charge distribution~\cite{shklovskii1999wigner,shklovskii1999screening}.
Lyubartsev et al~\cite{lyubartsev1997monte}, and Rouzina and Bloomfield~\cite{rouzina1996macroion} found that multivalent counterions 
bridging between macroions can also cause like-charge attractions. 
Netz and his coworkers ~\cite{Netz,moreira2001binding} later
developed a rigorous theoretical framework based on field theory to describe strong coupling correlations among counterions inside the condensation layer. 
This theory has since been applied to a wide variety of charged systems to describe like-charge attractions~\cite{ha1997counterion,linse1999electrostatic,vsamaj2011counterions,hatlo2010electrostatic,jho2006interaction,jho2008numerical,levin2002electrostatic,boroudjerdi2005statics,naji2013perspective,french2010long,kjellander1984correlation,kekicheff1993charge}. 

Most of the strong coupling (SC) theories based on the formalism by Netz consider point-like counterions whereas in most biological and soft matter systems, 
multivalent ions, such as spermidine, spermine~\cite{needleman2004higher}, often have elongated shapes ~\cite{alvarez1983large}. Sometimes, cylindrical 
molecules such as DNA, cytoskeletal filaments play the role of higher valence counterion. In the SC regime the multivalent counterions are 
usually very close to each other and the geometry of the charge distribution inside each counterion strongly influences their interactions. 
In macroions like polyelectrolytes additional mechanisms for like charge attractions, like inter-segment attraction forces and bridging forces, occur 
which is not possible in point charge systems \cite{israelachvili2011intermolecular, aakesson1989electric,podgornik2004polyelectrolyte}. The effects
of the geometry of the polyelectrolytes on their electrostatic properties have been studied before at the mean field level \cite{borukhov1999effect,may2008bridging}.  
Kim \textit{et. al.} \cite{kim2008attractions} were the first to introduce the finite size of the ions to the SC formalism to study the like-charge
attractions among two parallel plates in the presence of dumbbell-like shaped counterions. They found that in the SC regime the electrostatic potential 
appears to be flat and no energetic bridging occurs. Kanduc \textit{et al.} \cite{kanduvc2009role} later constructed a SC theory
for structured counterions. While their excluded volume interactions do not have any structure and are treated as point-like, their internal structures 
are described by the contributions from multipoles in the electrostatic interactions. They showed that in the SC regime the contributions of quadrupoles and higher order 
multipoles are important when there exist dielectric inhomogeneities.  When the dielectric constant is homogeneous over space, monopole contributions
are dominant in the SC regime. Bohinc and his coworkers \cite{bohinc2012interactions} developed an electrostatic theory of rodlike counterions confined between two plates, 
based on splitting the interactions into,
a short-ranged term calculated using cumulant expansion, and a long ranged term calculated using mean field. This ensures the validity of their theory
from the weak to the SC regimes. Using this formalism they could map out the region of attractions among the plates over a
wide range of the phase space. In most of these earlier models however the extended charges are treated as being composed of equi-spaced point 
particles.

In this work we treat the rodlike counterions from the viewpoint of uniformly charged polymers different from most of the 
discrete point-charge models studied before and we use polymer field theory to develop our formalism. The additional 
degrees of freedom in finite sized molecules add to the complexity of the many-body interactions. As a result in most of the SC literature 
on finite sized charges the calculations are limited to the zeroth order in the thermodynamic quantities. The use of the polymer fields allow us 
to systematically calculate terms beyond the zeroth order. In Section \ref{Sec1} we consider a system of polyelectrolytic counterions, 
the charges of which are uniformly smeared over the polymers, in the presence of a fixed charge distribution. The SC formalism is exact in 
the limit when the Coulomb coupling parameter $\Xi$ is very large or $1/\Xi$ is very small.
In the SC regime the counterions crowd near the oppositely charged surface, hence the interactions with the
surface at zeroth order dominates over the inter-particle interactions including the excluded volume interactions.
Thus the problem is effectively mapped to a single particle picture in the SC regime.  An alternate formalism of the strong coupling 
theory is based on the fact that the counterions form a Wigner crystal \cite{samajPRE,samajPRL} near the fixed charge surface and
predicts $1/\sqrt{\Xi}$ leading order term in the expansion of the thermodynamic quantities \cite{hatlo2010electrostatic}. The single particle picture would be valid when the Wigner lattice constant $b$ satisfies $b/\mu \propto \sqrt\Xi >> 1$, where $\mu $ is the characteristic
length of the system. In case of polyelectrolytic counterions of length $N$ an additional constraint $N < b$ should be satisfied for
the single particle picture to be valid. 
Here we closely follow the procedure of Netz \cite{Netz} and using polymer field theory obtain the expressions of the zeroth and the first order terms 
in the expansion in $1/\Xi$ of the grand partition function, density profile and free energy of the polymer counterions in the SC regime. 
Except for a few specific cases most polymer models are not analytically tractable because of the complicated functional integrals over the polymer fields.
However for rodlike polymers explicit forms of the single canonical partition function and the single polymer density which are central to the SC theory
are known in the literature \cite{fredrickson2006equilibrium}.
In Section \ref{Sec2} we work out the zeroth and first order terms of the density derived in Section \ref{Sec1} for the specific case of rodlike counterions
in the presence of a fixed charge distribution. We consider the special case when the fixed charge distribution is a charged wall and two charged walls
in Section \ref{subsectionA} and \ref{subsectionB} respectively. The dependence of the orientation averaged density profiles of the rods on their lengths and  
their excluded volume interactions with the walls are discussed. We also look at two different representations of the rods, one parameterized by the center 
of the rods and the other by one end of the rods and work out the thermodynamics in both case. We compare our results with the point-particle case of Moreira
and Netz \cite{moreira2001binding, Netz} 
to illustrate the effects of finite size of the rodlike ions on their thermodynamics. Comparison with the 
existing results of Kim \textit{et al.} \cite{kim2008attractions} and Bohinc \cite{bohinc2012interactions} are also made. In Section \ref{Sec3}
limitations of the present formalism and directions for future work are discussed.

\section{Thermodynamics of Polyelectrolytes in the SC limit}
\label{Sec1}
Consider a system of $n$ charged homo-polymer counterions in an external potential $-h(\mathbf{x})$.
Each polymer consists of $N$ statistical segments and charge $q$ smeared uniformly over each monomer.
There is also a fixed charged distribution with a surface charge density $-\sigma_s$, denoted 
by $-\rho_{c}(\mathbf{x})$. The system as a whole is charge neutral. All the charges interact via the Coulomb interactions $V(\vert\mathbf{x}-\mathbf{x}^{\prime}\vert) 
= 1/\vert\mathbf{x}-\mathbf{x}^{\prime}\vert$. The position in space of the segment $s$ of the polymer counterions is represented by a polymer field $\mathbf{r}(s)$.
The Hamiltonian of the system is given by
\begin{align}
 \beta H_n & = \sum_{i=1}^n\beta H_0^i +\frac{l_B}{2}\int d\mathbf{x}d\mathbf{x}^{\prime}\left[q\hat{\rho}(\mathbf{x})-\rho_{c}(\mathbf{x})\right]V(\mathbf{x}-\mathbf{x}^{\prime}) 
 \left[q\hat{\rho}(\mathbf{x}^{\prime})-\rho_{c}(\mathbf{x}^{\prime})\right] -\frac{l_B}{2}q^2nV_s \nonumber\\ 
  & - \int d\mathbf{x}h(\mathbf{x})\hat{\rho}(\mathbf{x}),
  \label{eq1.1}
\end{align}
where $H_0^i$ is the single polymer Hamiltonian, $\beta = 1/k_BT$ inverse temperature and $l_B = \beta e^2/\epsilon$ is the Bjerrum length. Here we consider the dielectric constant 
of the system $\epsilon$ to be 1. $H_0^i$ contains the 
information about the internal structure and bonding among the monomers and also 
the elastic properties like the stiffness or bending rigidity of the polymer counterions. The segment density function of the polymer 
counterions is defined by
\begin{equation}
 \hat{\rho}(\mathbf{x}) = \sum_{i}^{n}\int_{0}^{N}ds\delta(\mathbf{r}_{i}(s)-\mathbf{x}),
 \label{eq1.2}
\end{equation}
where $\mathbf{r}_{i}(s)$ denotes the polymer field for the $i$th polymer. We use the notation $\mathbf{x}$ 
for the position coordinate and $\mathbf{r}$ for the polymer field. $V_s$ is the self Coulomb interaction
between the different segments of the same polymer,
\begin{equation}
V_s = \int dsds^{\prime}V(\vert\mathbf{r}(s)-\mathbf{r}(s^{\prime})\vert).
 \label{eq1.3}
\end{equation}
Since we consider the polymer ions as being uniformly charged and the charges on the polymers are proportional to their lengths, 
it is not easy to go to the point-particle limit as in that limit 
the charge on the polymer goes to zero. Therefore we have to be cautious while taking that limit. 

We convert all the quantities in the above discussions to dimensionless units. All the position coordinates $\mathbf{x}$ 
and the polymer fields $\mathbf{r}$ are scaled by the Gouy-Chapman-like length scale $\mu = 1/(2\pi ql_B\sigma_s)$ , 
$\widetilde{\mathbf{x}}=\mathbf{x}/\mu$ and  $\widetilde{\mathbf{r}}=\mathbf{r}/\mu$.
Note the Gouy-Chapman-like length is defined in terms of the monomer charge $q$ not the total polymer charge. 
Hence the scaling length is same for all polymers irrespective of their lengths.
The strength of the interactions among the counterions is given by a dimensionless coupling parameter $\Xi = 2\pi q^3l_B^2\sigma_s$.
The dimensionless fixed charge distribution is defined by $\widetilde{\rho_{c}}(\mathbf{x}/\mu)
= \mu\rho_c(\mathbf{x})/\sigma_s$ and the dimensionless polymer density by $\hat{\rho}(\mathbf{x}/\mu) = \mu^3\hat{\rho}(\mathbf{x})$. 
In dimensionless units the Hamiltonian in equation \eqref{eq1.1} can be written as
\begin{align}
 \widetilde{H}_n & = \sum_{i=1}^n\widetilde{H}_0^i +\frac{\Xi}{2}\int d\widetilde{\mathbf{x}}d\widetilde{\mathbf{x}}^{\prime}
 \hat{\rho}(\widetilde{\mathbf{x}})\hat{\rho}(\widetilde{\mathbf{x}}^{\prime})\widetilde{V}(\vert\widetilde{\mathbf{x}}-\widetilde{\mathbf{x}}^{\prime}\vert)
 +\frac{1}{8\pi^2\Xi}\int d\widetilde{\mathbf{x}}d\widetilde{\mathbf{x}}^{\prime} 
 \widetilde{\rho}_c(\widetilde{\mathbf{x}})\widetilde{\rho}_c(\widetilde{\mathbf{x}}^{\prime})\widetilde{V}(\vert\widetilde{\mathbf{x}}-\widetilde{\mathbf{x}}^{\prime}\vert)
 \nonumber \\& + \int d\widetilde{\mathbf{x}}\widetilde{w}(\widetilde{\mathbf{x}})\hat{\rho}(\widetilde{\mathbf{x}})
  -\frac{\Xi}{2}\widetilde{V}_s - \int d\widetilde{\mathbf{x}}h(\widetilde{\mathbf{x}})\hat{\rho}(\widetilde{\mathbf{x}}).
  \label{eq1.4}
\end{align}
In the above Hamiltonian, we use the notation $\widetilde{w}(\widetilde{\mathbf{x}})$ for the electrostatic potential due to the fixed charge distribution on the polymer counterions 
\begin{equation}
\widetilde{w}(\widetilde{\mathbf{x}}) = -\frac{1}{2\pi}\int d\widetilde{\mathbf{x}}^{\prime}\widetilde{\rho}_c(\widetilde{\mathbf{x}}^{\prime})
\widetilde{V}(\vert\widetilde{\mathbf{x}}-\widetilde{\mathbf{x}}^{\prime}\vert)
 \label{eq1.5}
\end{equation}
Following Netz \cite{Netz} we add a vanishing term 
\begin{equation}
\frac{1}{4\pi^2\Xi}\int d\widetilde{\mathbf{x}}d\widetilde{\mathbf{x}}^{\prime} 
 \widetilde{\rho}_c(\widetilde{\mathbf{x}})\widetilde{\rho}_c(\widetilde{\mathbf{x}}^{\prime})\widetilde{V}(\vert\widetilde{\mathbf{x}}-\widetilde{\mathbf{x}}_0\vert)
 -\frac{n}{2\pi}\int d\widetilde{\mathbf{x}}^{\prime}\widetilde{\rho}_c(\widetilde{\mathbf{x}}^{\prime})\widetilde{V}(\vert\widetilde{\mathbf{x}}-\widetilde{\mathbf{x}}_0\vert) = 0
 \label{eq1.6}
\end{equation}
to the Hamiltonian that cancels the divergences that results from the long range nature of the Coulomb interactions. This term vanishes because of the charge neutrality condition
$\int d\widetilde{\mathbf{x}}\widetilde{\rho}_c(\widetilde{\mathbf{x}}) = 2\pi\Xi n $. The coordinate point $\widetilde{\mathbf{x}}_0$ in the above equation is specified in the next Section.
After adding the vanishing term in equation \eqref{eq1.6}, the Hamiltonian reads
\begin{align}
 \widetilde{H}_n & = \sum_{i=1}^n\widetilde{H}_0^i +\frac{\Xi}{2}\int d\widetilde{\mathbf{x}}d\widetilde{\mathbf{x}}^{\prime}
 \hat{\rho}(\widetilde{\mathbf{x}})\hat{\rho}(\widetilde{\mathbf{x}}^{\prime})\widetilde{V}(\vert\widetilde{\mathbf{x}}-\widetilde{\mathbf{x}}^{\prime}\vert)
 +\frac{1}{4\pi^2\Xi}\int d\widetilde{\mathbf{x}}d\widetilde{\mathbf{x}}^{\prime} 
 \widetilde{\rho}_c(\widetilde{\mathbf{x}})\widetilde{\rho}_c(\widetilde{\mathbf{x}}^{\prime})\biggl[\widetilde{V}(\vert\widetilde{\mathbf{x}}-\widetilde{\mathbf{x}}^{\prime}\vert)/2
 \nonumber \\& -\widetilde{V}(\vert\widetilde{\mathbf{x}}-\widetilde{\mathbf{x}}_0\vert)\biggr] + \int d\widetilde{\mathbf{x}}\widetilde{w}(\widetilde{\mathbf{x}})\hat{\rho}(\widetilde{\mathbf{x}})
  -\frac{\Xi}{2}n\widetilde{V}_s - \int d\widetilde{\mathbf{x}}h(\widetilde{\mathbf{x}})\hat{\rho}(\widetilde{\mathbf{x}}),
  \label{eq1.7}
\end{align}
where we have redefined the potential due to the fixed charge
\begin{equation}
\widetilde{w}(\widetilde{\mathbf{x}}) = -\frac{1}{2\pi}\int d\widetilde{\mathbf{x}}^{\prime}\widetilde{\rho}_c(\widetilde{\mathbf{x}}^{\prime})
\biggl[\widetilde{V}(\vert\widetilde{\mathbf{x}}-\widetilde{\mathbf{x}}^{\prime}\vert)- \widetilde{V}(\vert\widetilde{\mathbf{x}}^{\prime}-\widetilde{\mathbf{x}}_0\vert)\biggr].
 \label{eq1.8}
\end{equation}

The grand partition function of the system with the above Hamiltonian is 
\begin{equation}
  \mathcal{Z} = \sum\limits_{n=0}^{\infty}\frac{\lambda^n\mu^{3n}}{n!\lambda_T^{3n}}\int\left[\prod\limits_{i=1}^{n}\mathcal{D}\widetilde{\mathbf{r}}_i
  \widetilde{\Omega}(\widetilde{\mathbf{r}}_i)\right]\exp(-\widetilde{H}_n[\{\widetilde{\mathbf{r}}_i\}]),
  \label{eq1.9}
\end{equation}
where $\lambda$ denotes the fugacity, $\lambda_T = \sqrt{h^2\beta/2\pi m}$ the thermal wavelength and $\mathcal{D}$ the integral over the polymer field 
configurations. The square brackets represents the functional dependence of $\widetilde{H}_n$ on the polymer fields $\widetilde{\mathbf{r}}(s)$. 
The term $\mu^{3n}$ comes from the rescaling of the polymer fields $\widetilde{\mathbf{r}}_i$. The function $\widetilde{\Omega}(\widetilde{\mathbf{x}})$
represents the amount of available space
to the counterions which might be confined by hard walls or traps. Since the SC theory is exact in the limit when $1/\Xi\rightarrow 0$,
we introduce a scaled fugacity defined by $\Lambda =2\pi\mu^3\lambda\Xi/\lambda_T^3$ that enable us to rewrite equation \eqref{eq1.9} in a series in $1/\Xi$
\begin{equation}
  \mathcal{Z} = \sum\limits_{n=0}^{\infty}\left(\frac{\Lambda}{2\pi\Xi}\right)^nQ_n.
  \label{eq1.10}
\end{equation}
$Q_n$ is the canonical parition functions for $n$ polymers and has the functional form
\begin{align}
Q_n[h-\widetilde{w},\widetilde{V}]  & = \frac{\mathcal{Z}_0}{n!}\int\left[\prod\limits_{i=1}^{n}\mathcal{D}\widetilde{\mathbf{r}}_i
  \widetilde{\Omega}(\widetilde{\mathbf{r}}_i)\right]\exp\biggl(-\sum\limits_{i=1}^nH_0[\widetilde{\mathbf{r}_i}]-\Xi\sum_{i<j}\int_{0}^{N}ds
  \int_{0}^{N}ds^{\prime}V(\vert\widetilde{\mathbf{r}_i}(s)- \widetilde{\mathbf{r}_j}(s^{\prime})\vert) \nonumber\\ & - \sum\limits_{i=1}^n\int_0^Nds
  \widetilde{w}[\widetilde{\mathbf{r}_i}(s)]+ \sum\limits_{i=1}^n\int_0^Ndsh[\widetilde{\mathbf{r}_i}(s)]\biggr)
 \label{eq1.11}
\end{align}
where 
\begin{align}
\mathcal{Z}_0  = \exp\left(-\frac{1}{4\pi^2\Xi}\int d\widetilde{\mathbf{x}}d\widetilde{\mathbf{x}}^{\prime}\widetilde{\rho}_c(\widetilde{\mathbf{x}})
\widetilde{\rho}_c(\widetilde{\mathbf{x}}^{\prime})\biggl[\widetilde{V}(\vert\widetilde{\mathbf{x}}-\widetilde{\mathbf{x}}^{\prime}\vert)/2
   -\widetilde{V}(\vert\widetilde{\mathbf{x}}-\widetilde{\mathbf{x}}_0\vert)\biggr]\right).
 \label{eq1.12}
\end{align}
From equation \eqref{eq1.10} we see that in the SC limit only the lowest order canonical
partition functions contributes the most to the grand partition function.
An alternate expression for the grand partition function is obtained by the Hubbard-Stratonovich \cite{hubbard1959calculation,
stratonovich1957method} transformation 
\begin{align}
 \mathcal{Z} & = \mathcal{Z}_0\sum_{n=0}^{\infty}\frac{1}{n!}\left(\frac{\Lambda}{2\pi\Xi}\right)^n
 \biggl\langle \biggl(Q_1[h - \widetilde{w} - i\phi]\biggr)^n\biggr\rangle_{\phi},
 \label{eq1.13}
\end{align}
where $Q_1$ is a single polymer partition function defined in equation \eqref{eq1.11} and $\phi$ is the field introduced during the transformation \cite{kanduvc2009role, Netz}.
The averaging $\langle..\rangle_{\phi}$ is performed with respect to the weight factor 
\begin{equation}
\frac{1}{2\Xi}\int_{\widetilde{\mathbf{x}},\widetilde{\mathbf{x}^{\prime}}}
\phi(\widetilde{\mathbf{x}})V^{-1}(\vert\widetilde{\mathbf{x}} -\widetilde{\mathbf{x}^{\prime}}\vert)\phi(\widetilde{\mathbf{x}^{\prime}}).
\label{eq1.14}
\end{equation}
The derivation of equation \eqref{eq1.13} is similar to the point particle case \cite{Netz}. The details are shown in Appendix \ref{appendix0}. Equations \eqref{eq1.13} can be shown
to be equivalent to \eqref{eq1.10} by performing the Gaussian averaging explicitly. The point particle case is obtained by using $Q_1[i\phi] = \int 
d\widetilde{\mathbf{x}}e^{-i\phi(\widetilde{\mathbf{x}})}$.

The free energy has a similar expansion in $1/\Xi$ in the SC limit
\begin{align}
 \mathcal{F}_n  = \frac{\mathcal{F}_1}{\Xi} + \frac{\mathcal{F}_2}{\Xi^2} + \mathcal{O}(1/\Xi^3).
 \label{eq1.15}
\end{align}
The detailed forms of the first and second order terms $\mathcal{F}_1$ and $\mathcal{F}_2$ are worked out in Appendix \ref{appendixB}. 
 We define the rescaled density distribution as $\widetilde{\rho}(\widetilde{\mathbf{x}})=2\pi\Xi\delta\ln\mathcal{Z}/\delta h(\widetilde{\mathbf{x}})$.
 Similarly the density has an expansion in $1/\Xi$ as 
\begin{align}
 \widetilde{\rho}(\widetilde{\mathbf{x}})=\widetilde{\rho}_0(\widetilde{\mathbf{x}})+\frac{1}{\Xi}\widetilde{\rho}_1(\widetilde{\mathbf{x}}) +
 \mathcal{O}(1/\Xi^2).
 \label{eq1.16}
\end{align}
The zeroth order term in the density is derived from equations \eqref{eq1.13} and \eqref{eq1.16} of the partition function   
\begin{align}
 \widetilde{\rho}_0(\widetilde{\mathbf{x}}) = \Lambda\left\langle\frac{\delta Q_1[ h - \widetilde{w} -i\phi]}{\delta h(\widetilde{\mathbf{x}})}
 \right\rangle_{\phi}\bigg\vert_{h = 0}. 
 \label{eq1.17}
 \end{align}
 with the averaging with respect to the weight in equation \eqref{eq1.14}. If we denote the single polymer density
 operator in an external field $\phi$ by $\rho^{(1)}(\widetilde{\mathbf{x}};\phi) = \delta Q_1[\phi]/\delta\phi(\widetilde{\mathbf{x}})$ 
 \cite{fredrickson2006equilibrium}, the above equation can be rewritten as
 \begin{align}
 \widetilde{\rho}_0(\widetilde{\mathbf{x}}) = \Lambda\left\langle\rho^{(1)}(\widetilde{\mathbf{x}};\widetilde{w}+i\phi)\right\rangle_{\phi}. 
 \label{eq1.18}
 \end{align}
 Alternatively from equation \eqref{eq1.10} the zeroth order density term is calculated to be
 \begin{align}
 \widetilde{\rho}_0(\widetilde{\mathbf{x}}) & = \frac{\delta Q_1[ h - \widetilde{w}]}{\delta h(\widetilde{\mathbf{x}})} \biggl\vert_{h=0}\nonumber \\
 & = \rho^{(1)}(\widetilde{\mathbf{x}};\widetilde{w}).
 \label{eq1.19}
 \end{align}
 Similarly the first order term in the density distribution function has the form
\begin{align}
 \widetilde{\rho}_1(\widetilde{\mathbf{x}}) & = \frac{\Lambda^2}{2\pi}\left(\left\langle Q_1\frac{\delta Q_1}{\delta h(\widetilde{\mathbf{x}})}
 \right\rangle_{\phi} 
 - \left\langle Q_1\right\rangle_{\phi}\left\langle \frac{\delta Q_1}{\delta h(\widetilde{\mathbf{x}})}\right\rangle_{\phi}\right)\bigg\vert_{h=0}.
 \label{eq1.20}
 \end{align}
 In terms of the single polymer density operator $\rho^{(1)}(\widetilde{\mathbf{x}};\phi)$ the above equation becomes
 \begin{align}
 \widetilde{\rho}_1(\widetilde{\mathbf{x}}) & = \frac{\Lambda^2}{2\pi}\left(\left\langle Q_1[ \widetilde{w} +i\phi]\rho^{(1)}(\widetilde{\mathbf{x}};
  \widetilde{w} +i\phi)\right\rangle_{\phi} - \left\langle Q_1[ \widetilde{w} +i\phi]\right\rangle_{\phi}\left\langle\rho^{(1)}(\widetilde{\mathbf{x}};
   \widetilde{w} +i\phi)\right\rangle_{\phi}\right).
 \label{eq1.21}
 \end{align}
We get another expression for the leading density term from equation \eqref{eq1.10}
\begin{align}
 \widetilde{\rho}_1(\widetilde{\mathbf{x}}) & = \frac{\delta (Q_2[ h - \widetilde{w},\widetilde{V}]-Q_1^2[ h - \widetilde{w}])}{\delta h(\widetilde{\mathbf{x}})} 
 \biggl\vert_{h=0}.
 \label{eq1.22}
 \end{align}
Because of two equivalent representations of the grand partition function in equations \eqref{eq1.10} and \eqref{eq1.13}, we have two alternate 
representations of the zeroth (equations \eqref{eq1.18} and \eqref{eq1.19}) and the first order (equations \eqref{eq1.21} and \eqref{eq1.22}) density distributions too.
We use either representation depending on the ease of calculations as shown in Appendix \ref{appendixC}.

\section{Applications: Rodlike polymer counterions}
\label{Sec2}

For most polymer models it is not possible to obtain analytical expressions for the thermodynamic quantities derived in the previous Section
and are usually computed numerically. However for rodlike polymers the single particle canonical partition function $Q_1$ which is central to
the discussions above, has a nice analytical form that allows us to obtain explicit expressions for the thermodynamic quantities.
The rods depend on an additional orientation variable $\mathbf{u}$ that determines the alignment of the rods. 
For rodlike polymers the single chain partition function $Q_1$ in an external potential $\widetilde{w}$ is given by \cite{fredrickson2006equilibrium}
\begin{equation}
Q_1[\widetilde{w}] = \frac{1}{4\pi}\int_{\Omega} d\mathbf{\widetilde{x}}\int d\mathbf{u}\exp\left(-\int_{0}^{N}ds\widetilde{w}(\mathbf{\widetilde{x}}+s\mathbf{u})\right),
 \label{eq2.0}
\end{equation}
where we have used $\int_{\Omega}$ to denote the integral $\int \widetilde{\Omega}(\widetilde{\mathbf{x}},\mathbf{u})$ over the available space. From this equation we 
get the single polymer density operator to be \cite{fredrickson2006equilibrium}
 \begin{align}
 \rho^{(1)}(\widetilde{\mathbf{x}},\mathbf{u}) = \frac{\Lambda}{4\pi}\int_{0}^{N} ds \exp\left[-\int_{0}^{N}ds^{\prime}\widetilde{w}(\widetilde{\mathbf{x}}+(s^{\prime}-s)\mathbf{u})
 \right].
 \label{eq2.1}
\end{align}
From equations \eqref{eq2.0} and \eqref{eq2.1} the zeroth order density is computed using
equation \eqref{eq1.19} and the first order after Gaussian averaging of equation \eqref{eq1.20}. The detailed calculations of the density terms are 
worked out in Appendix \ref{appendixC}. Here we quote the final results
\begin{align}
  \widetilde{\rho_0}(\widetilde{\mathbf{x}},\mathbf{u}) & = \Lambda \int_{0}^{N} ds q_2(\widetilde{\mathbf{x}},\mathbf{u},s), \label{eq2.2}\\
 \widetilde{\rho_1}(\widetilde{\mathbf{x}},\mathbf{u}) & = \frac{\Lambda^2}{2\pi}\int_{\Omega} d\widetilde{\mathbf{x}}^{\prime}\int d\mathbf{u}^{\prime}\int_{0}^{N} ds 
 q_1(\widetilde{\mathbf{x}},\mathbf{u}^{\prime})q_2(\widetilde{\mathbf{x}},\mathbf{u},s)\biggl[\exp\biggl(-\Xi\int ds^{\prime}ds^{\prime\prime}
  V\biggl(\vert\widetilde{\mathbf{x}}+(s^{\prime}-s)\mathbf{u}- \nonumber\\& \widetilde{\mathbf{x}}^{\prime} - s^{\prime\prime}\mathbf{u}^{\prime}\vert\biggr)\biggr)-1\biggr], 
  \label{eq2.3}
  \end{align}
with the quantities 
\begin{align}
 q_1(\widetilde{\mathbf{x}}, \mathbf{u}) & = \frac{1}{4\pi}\exp\biggl(-\int_{0}^{N}ds\widetilde{w}(\widetilde{\mathbf{x}}+s\mathbf{u})\biggr), \label{eq2.4}\\
 q_2(\widetilde{\mathbf{x}}, \mathbf{u}, s) & = \frac{1}{4\pi}\exp\biggl(-\int_{0}^{N}ds^{\prime}\widetilde{w}(\widetilde{\mathbf{x}}+(s^{\prime}-s)\mathbf{u})\biggr).
 \label{eq2.5}
\end{align}

To demonstrate the formalism we work out two specific cases: In Subsection \ref{subsectionA} we derive the thermodynamics of rodlike counterions confined in 
a half plane by a charged wall and in Subsection \ref{subsectionB} we study the rodlike counterions confined between two charged walls. The walls assumed to be 
impenetrable hence we explicitly consider excluded volume interactions between the rods and the wall through $\widetilde{\Omega}(\widetilde{\mathbf{x}},\mathbf{u})$,
however to keep things simple 
we neglect the steric interactions among the rods themselves. We compare our results with that of the
point-particles, the dumbbell-like counterions of Kim \textit{et al.} \cite{kim2008attractions} and the rodlike polymers of Bohinc \cite{bohinc2012interactions}.

\subsection{One charged wall}
\label{subsectionA}

\begin{figure*}[h]
        \centering
           \subfloat{%
              \includegraphics[width=8cm,trim={1cm 5cm 0 2cm},clip]{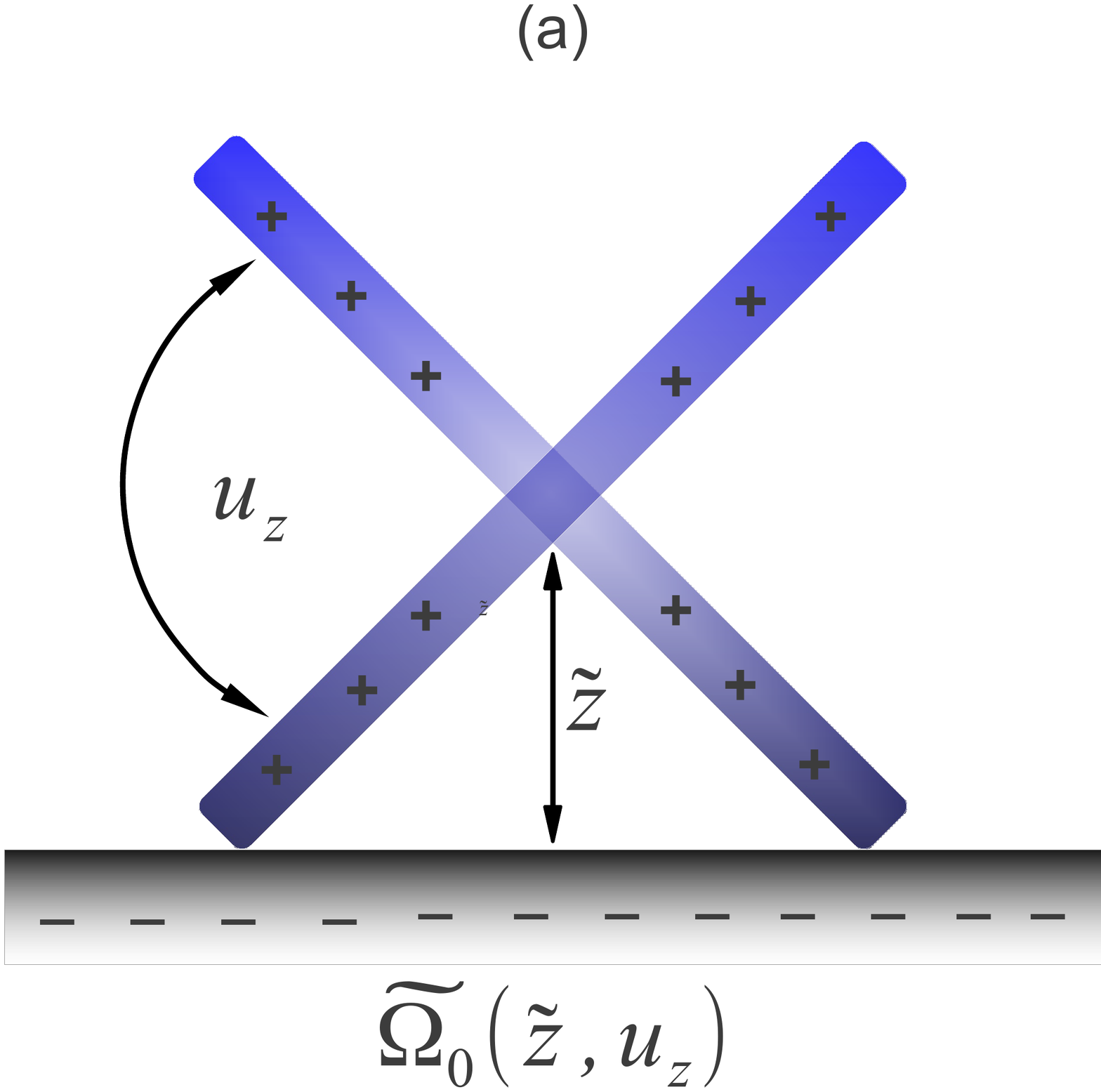}%
           } 
           \subfloat{%
              \includegraphics[width=8cm,trim={0cm 4.8cm 0 2cm},clip]{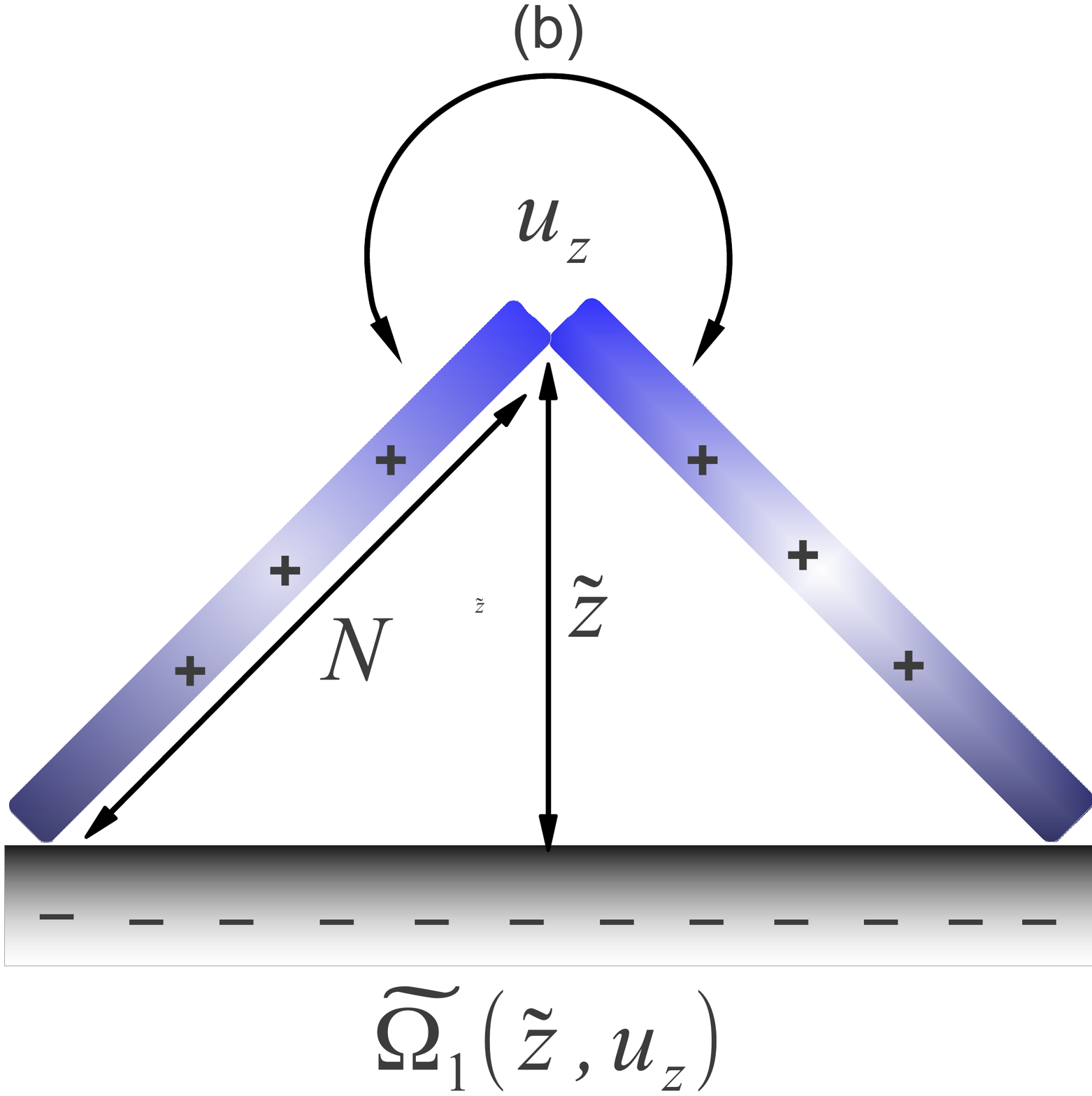}%
           }
           \caption{(Color online) Schematic diagram showing the volume available to the charged rods close to the oppositely charged wall. 
           The rods can be parameterized in two ways: when $\widetilde{z}$
           represents (a) the center of the rods and (b) one end of the rods. Accordingly the available volume to a rod, $\widetilde{\Omega}(\widetilde{z}, u_z)$
           in equation \eqref{eq1.9}, has two different functional forms: (a) $\widetilde{\Omega}_0(\widetilde{z}, u_z)$
           in equation \eqref{eq2A.3} and (b) $\widetilde{\Omega}_1(\widetilde{z}, u_z)$ in equation \eqref{eq2A.9}.}
           \label{figomega}
 \end{figure*}
 
We consider a system of rodlike counterions confined in the upper $z$ plane by a charged plate located at $z = 0$. The rescaled charge distribution of the plate is then,
$\widetilde{\rho_0}(\widetilde{\mathbf{x}}) = \delta(\widetilde{z})$. The scaled potential in this case is 
\begin{equation}
 \widetilde{w}(\widetilde{z}) = \widetilde{z},
 \label{eq2A.1}
\end{equation}
where the reference point $\widetilde{\mathbf{x}_0}$ in equation \eqref{eq1.11} is chosen to lie on the charged plate \cite{Netz}. The thermodynamics of the
system depends only on $\widetilde{z}$ coordinate and is effectively an 1D system. The orientation variable also depend only on $u_z$ ( the $z$ component 
of the orientation vector $\mathbf{u}$) given by $u_z = \cos\theta$, where $\theta$ is 
the angle between the the rods and the positive $z$ axis.  Because of the finite size of the rodlike polymers there are two ways to parameterize
them, one by choosing $\widetilde{z}$ at the center of the rods as the reference point and another by choosing $\widetilde{z}$ at one end of the rods. 
The two different parameterizations are shown in Figure \ref{figomega}. In the previous Section, $\widetilde{z}$ is located at one end of the rods which corresponds to the second 
parameterization. In the first case, $\widetilde{z}$ would denote the position of the center of the rods but then the segment integrals in equations 
\eqref{eq2.0}-\eqref{eq2.5} are replaced by $\int_{-N/2}^{N/2}$. We demonstrate both the parameterizations in the rest of this Section.

First we look at the parameterization with the center of the rods. From the schematic diagram Figure \ref{figomega}-(a) we see that the available volume function 
should be defined as
\begin{align}
 \widetilde{\Omega}_0(\widetilde{z},u_z) & = 1 \hspace{2mm}\text{for $0<\widetilde{z} < N/2$ and $-2\widetilde{z}/N<u_z<2\widetilde{z}/N$}, \nonumber\\
 & = 1 \hspace{5 mm} \text{$\widetilde{z} > N/2$ and} \nonumber\\
 & = 0 \hspace{5mm}\text{otherwise}.
 \label{eq2A.3}
\end{align}
Plugging the wall interactions, $\widetilde{w}(\widetilde{z}) = \widetilde{z}$,  into equation \eqref{eq2.2} the zeroth order term of the density profile 
becomes 
\begin{align}
 \widetilde{\rho}_0(\widetilde{z}, u_z) &  = \frac{1}{2}\Lambda\widetilde{\Omega}_0(\widetilde{z},u_z)\int_{-N/2}^{N/2} ds \exp\left[-N(\widetilde{z}-su_z)\right] \nonumber\\
 & =  \Lambda\widetilde{\Omega}_0(\widetilde{z},u_z)\frac{\sinh(N^2u_z/2)}{Nu_z}e^{-Nz}.
 \label{eq2A.2}
\end{align}

To obtain the leading order density term in equation \eqref{eq2.3} the fugacity $\Lambda$ need to be calculated. In the SC
regime the fugacity has an expansion in $1/\Xi$ as
\begin{equation}
 \Lambda = \Lambda_0 + \Lambda_1/\Xi + \mathcal{O}(1/\Xi^2).
 \label{eq2A.3.1}
\end{equation}
The first two terms of the fugacity is determined from the charge normalization condition 
\begin{align}
 \int d\widetilde{z}\widetilde{\rho}(\widetilde{z}) = 1.
 \label{eq2A.4}
\end{align}
The details of the procedure are given in Appendix \ref{appendixA} where we obtain the explicit expressions for the fugacity terms. The leading order contribution to the density 
distribution in equation \eqref{eq2.3} diverges due to the long ranged form of the Coulomb interactions, however this divergence is canceled by a 
corresponding diverging term in $\Lambda_1$ coming from the zeroth order density term in equation \eqref{eq2.2}. The first order density term gets renormalized and 
stays finite. The algebra is shown in Appendix \ref{appendixD}, equations \eqref{eqD.7}-\eqref{eqD.10}, here we write the final result  
\begin{align}
 \widetilde{\rho_1}(\widetilde{\mathbf{x}},\mathbf{u}) & = \frac{1}{N}\Lambda_0^2\int dsq_2(\widetilde{z},u_z,s)\int ds^{\prime}ds^{\prime\prime}
 \int d\mathbf{u}^{\prime}q_1(0,u_z^{\prime})\biggl[\frac{1}{2}\left(\widetilde{z}_s^2 - \langle(\widetilde{z}_s)^2\rangle\right) - 
 \frac{1}{N}\left(\widetilde{z}_s - \langle\widetilde{z}_s\rangle\right)\biggr].
 \label{eq2A.5}
\end{align}
$\widetilde{z}_s$  is a function of $\widetilde{z}$, $u_z$ and $s$
\begin{align}
\widetilde{z}_s(\widetilde{z},u_z,u_z^{\prime},s,s^{\prime},s^{\prime\prime})  =  \widetilde{z} + (s^{\prime}-s)u_z - s^{\prime\prime}u^{\prime}_z.   
\label{eq2A.6}
\end{align}
This density term contains the complex many-body effects of the extended geometry of the rodlike counterions and some approximations have been made to obtain
the above expression. However it captures all the essential effects of the finite size of the counterions. Since in the point particle
limit the orientational average is zero, $\langle z_s\rangle = 0$ and for $N = 1$ we recover the point-particle density in the SC limit.

We can also average over the orientations of the rods to obtain an orientation averaged density. We use the same notation $\widetilde{\rho}_0(\widetilde{z})$ for the
orientation averaged density. However we have to be careful to properly 
define the measure of the orientation term because of the nature of the excluded volume term $\widetilde{\Omega}_0$. A simple orientation average 
 $ \int_{-2\widetilde{z}/N}^{2\widetilde{z}/N}du_z \widetilde{\rho}_0(\widetilde{z},u_z) $
, obtained from the definition \eqref{eq2A.2}, vanishes as the center of the rods approaches the wall. Thus at the wall we get, $\widetilde{\rho}_0(\widetilde{z}) = 0$
as this measure can not count the number the rods aligned parallel to the plane $u_z = 0$. This is an artifact of using a continuous instead of a discrete measure. 
So we define the integral measure over the orientations in the following way to include the $u_z = 0$ orientation of the rods 
\begin{equation}
 \widetilde{\rho}_0(\widetilde{z}) = \int_{-2\widetilde{z}/N}^{0}du_z \widetilde{\rho}_0(\widetilde{z},u_z) + \widetilde{\rho}_0(\widetilde{z},u_z = 0) +
 \int_{0}^{2\widetilde{z}/N}du_z \widetilde{\rho}_0(\widetilde{z},u_z).
 \label{eq2A.7}
\end{equation}
With this definition we get the orientation averaged density as
\begin{align}
 \widetilde{\rho_0}(\widetilde{z}) & = \frac{2\Lambda_0}{N}\left(N^2/4 + \sinhInt(N\widetilde{z})\right)e^{-N\widetilde{z}} \hspace{2mm}\text{for $0<\widetilde{z}<N/2$ and}\nonumber\\
 & = \frac{2\Lambda_0}{N}\left(N^2/4 + \sinhInt(N^2/2)\right)e^{-Nz}\hspace{2mm}\text{for $\widetilde{z} > N/2$},
 \label{eq2A.8}
\end{align}
where 
\begin{equation}
\sinhInt(y) = \int_0^y dt\sinh(t)/t. 
 \label{eq2A.85}
\end{equation}

\begin{figure*}[h]
        \centering
           \subfloat{%
              \includegraphics[scale=0.4]{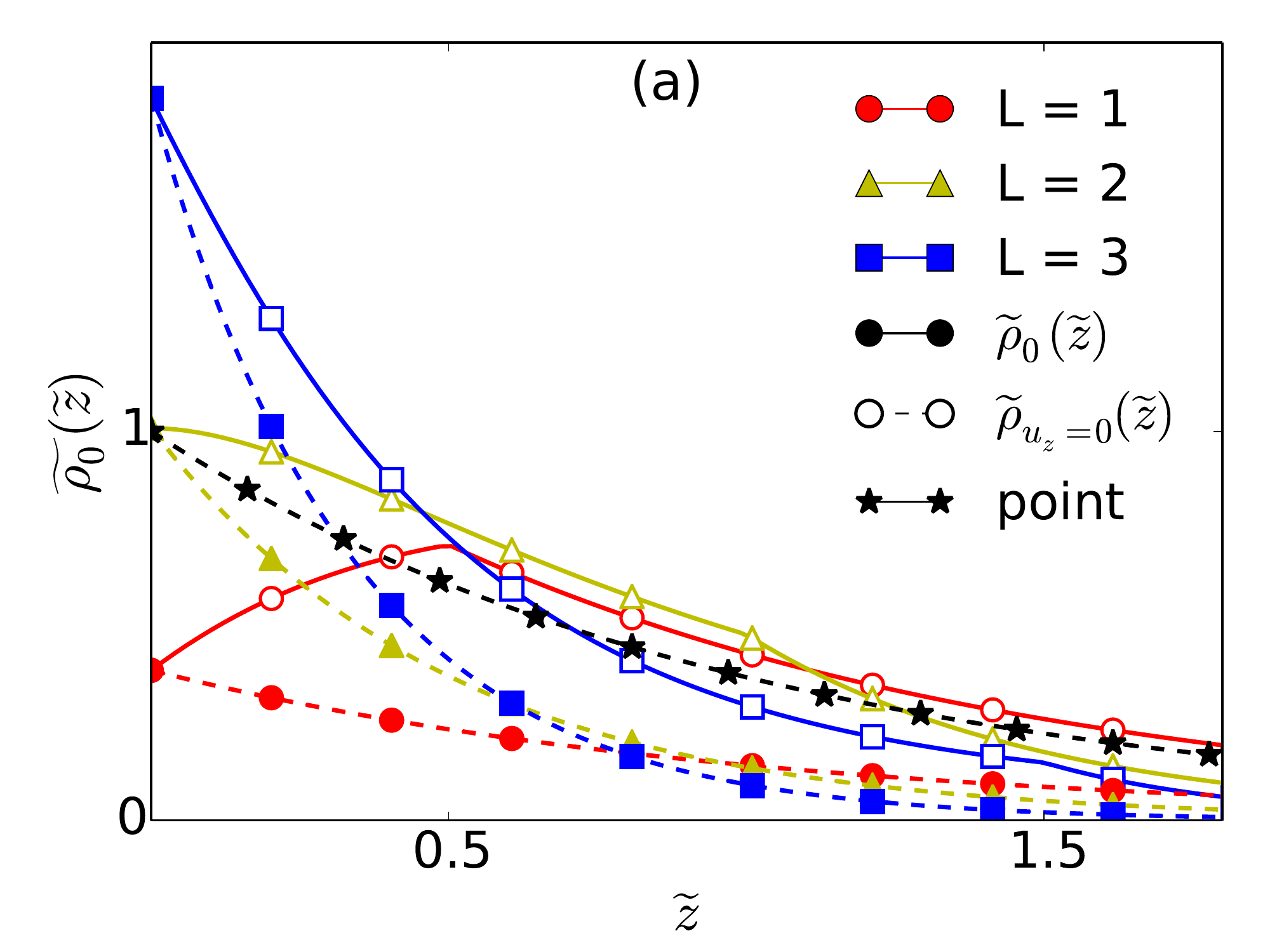}%
           } 
           \subfloat{%
              \includegraphics[scale=0.4]{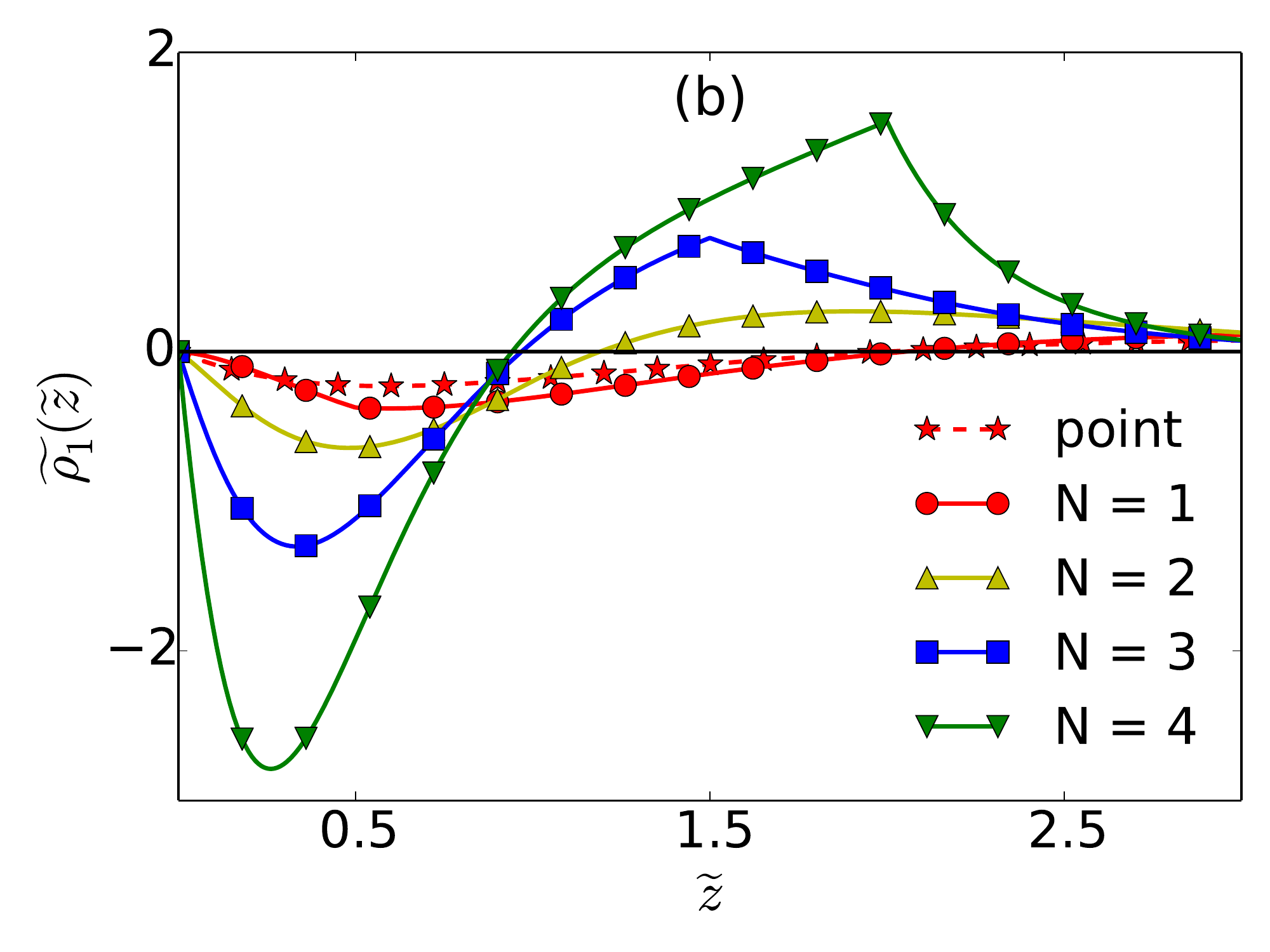}%
           }
           \caption{(Color online) One wall: (a) Zeroth order orientation averaged density (solid) for the rods in the SC limit according to equation \eqref{eq2A.8} and density of rods aligned parallel
           to the wall ( obtained from equation \eqref{eq2A.2} by using $u_z = 0$ ) denoted by $\rho_{u_z=0}$ (dotted) for various length of the rodlike counterions. Here $\widetilde{z}$ is the position of the center
           of the rods. Also shown is the point-particle density. 
           (b) Leading order density term for different lengths of the rods and the corresponding term for the point-particle case.}
           \label{Fig1}
 \end{figure*}
 
From our definition of the orientation averaged density and Figure \ref{Fig1}-(a), we see that all the counterions are aligned parallel to the plane very close to wall because of the excluded volume effects.
However as we go further away from the wall the counterions can rotate freely and the contribution to the density from the other orientational configurations becomes important.
This restriction of the orientations of the rods by the wall causes depletion of the rod densities near the wall. Smaller rods are much less affected by the excluded volume constraint 
than the larger ones, hence the other orientational configurations contribute much more to the density. The electrostatic correlations among the longer rods and the wall are much stronger
and hence they are more crowded near the wall. Figure \ref{Fig1}-(b) shows the first order density term for various lengths of the rodlike counterions. Since the leading order density term 
contains the information on the rod-rod interactions, Figure \ref{Fig1}-(b) says that longer rods are more highly correlated than the short ones. From both Figure \ref{Fig1}-(a) and Figure \ref{Fig1}-(b) 
we see that smaller rods behave similar to the point-particles as expected. 

\begin{figure*}[h]
        \centering
           \subfloat{%
              \includegraphics[scale=0.5]{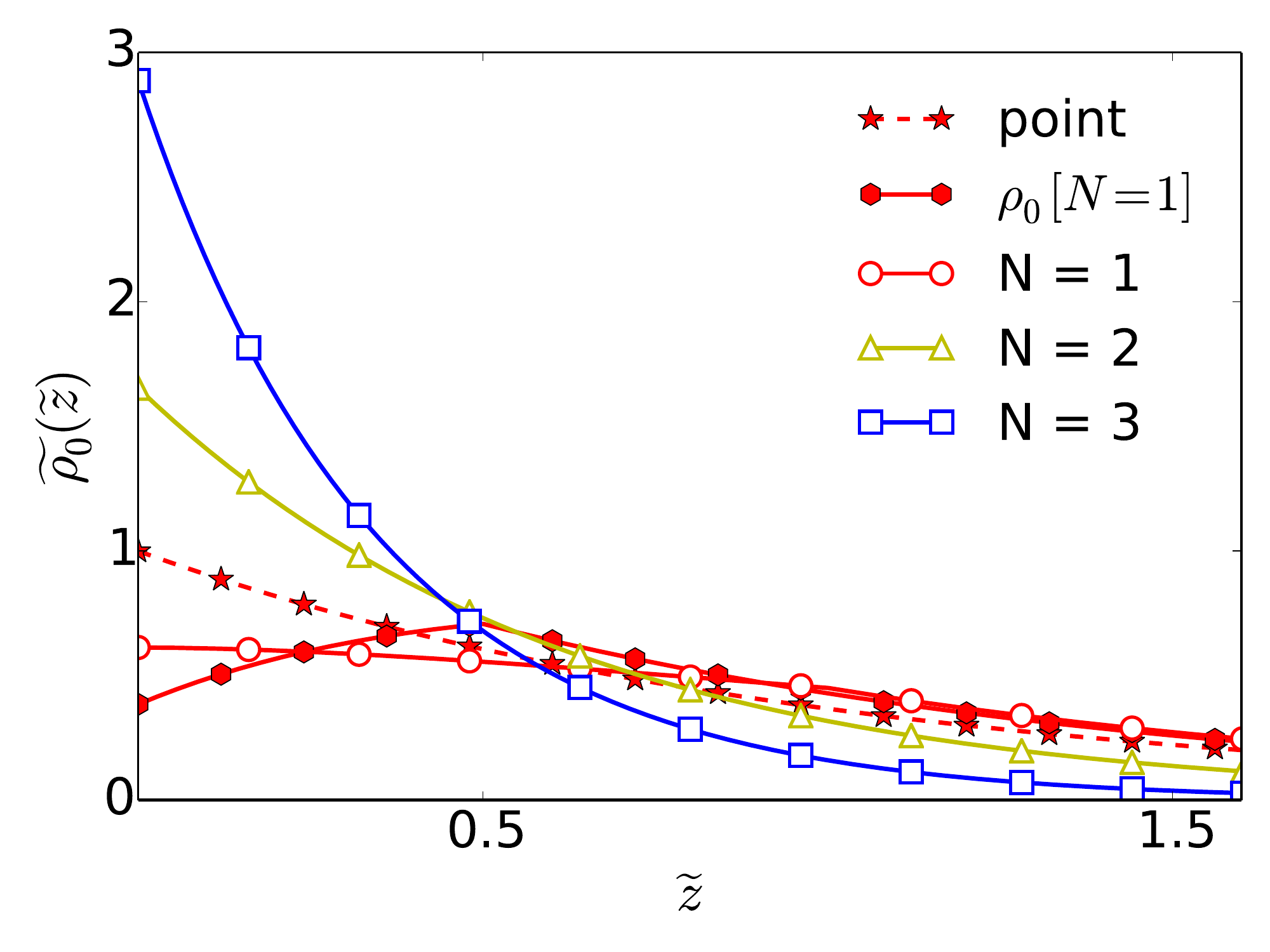}%
           } 
%            \subfloat{%
%               \includegraphics[scale=0.4]{plots/1/free_energy_pincus.pdf}%
%               \label{Fig3b}%
%            }
           \caption{(Color online) One wall: Zeroth order density for the rods calculated from equation \eqref{eq2A.10} when $\widetilde{z}$ represents one end of the rod. 
           The point-particle case and the density for the center of rod parameterization in equation \eqref{eq2A.8} is also shown.}
%            (b) Free energy of the counterions from equation \eqref{eqB.8} as a function of the rod length.}
           \label{Fig3}
 \end{figure*}

Note that there is an alternative analysis by Kim \textit{et al.} \cite{kim2008attractions}, where they define $\widetilde{z}$ at one end of the rods. 
Based on Figure \ref{figomega}-(b) we have to redefine the available volume term as
\begin{align}
 \widetilde{\Omega}_1(\widetilde{z},u_z) & = 1 \hspace{2mm}\text{for $0<\widetilde{z} < N$ and $-\widetilde{z}/N < u_z < 1$,} \nonumber\\
 & = 1 \hspace{5 mm} \text{$\widetilde{z} > N$ and } \nonumber\\
 & = 0 \hspace{5mm}\text{otherwise.}
 \label{eq2A.9}
\end{align}
With this new weight (measure) we obtain the zeroth order orientation averaged density as
\begin{align}
 \widetilde{\rho_0}(\widetilde{z}) & = \frac{\Lambda_0}{N}\left(\sinhInt(N^2/2) + \sinhInt(N\widetilde{z}/2)\right)e^{-N\widetilde{z}}
 \hspace{2mm}\text{for $0<\widetilde{z}<N$},\nonumber\\
 & = \frac{2\Lambda_0}{N}\sinhInt(N^2/2)e^{-Nz}\hspace{2mm}\text{for $\widetilde{z} > N$},
\label{eq2A.10}
 \end{align}
with $\Lambda_0 = 2N^2/(\expIntEi(-N^2/2)-\expIntEi(-3N^2/2)+\ln(3)+2\sinhInt(N^2/2))$ and $\expIntEi(y) = \int_{-y}^{\infty}dt\exp(-t)/t$. 
The zeroth order orientation averaged density is plotted in Figure \ref{Fig3}.
The general behavior of the polymer densities are qualitatively same in both the parameterizations as seen from Figures \ref{Fig1} and \ref{Fig3}.
Figure \ref{Fig3} shows quantitative agreement for short polymers for the two parameterizations as the midpoint and the endpoint of the rods 
coincide in this case. 
% We also plot the zeroth order term of the free energy for the counterions only in Figure \ref{Fig3b}, calculated using the formula \eqref{eqB.7}. Short rods have 
% smaller charges and hence less repulsion energy and more orientational degrees which cause them to lower their free energies than the longer ones. 

\subsection{Two charged wall}
\label{subsectionB}
When the rodlike counterions are confined between two charged plates separated by a distance $\widetilde{d}$, the electrostatic potential vanishes, 
$\widetilde{w}(\widetilde{z}) = 0$ \cite{Netz}.
Thus the two functions $q_1(\widetilde{z},u_z) = q_2(\widetilde{z},u_z,s) = \frac{1}{2}$ are constants. We first consider the parameterization with the center of the rods.
Even though $\widetilde{w}(\widetilde{z}) = 0$ the density of the rods still has a spatial  dependence due to the excluded volume term $\widetilde{\Omega}_0$
\begin{align}
 \widetilde{\rho}_0(\widetilde{z}, u_z) & =  \frac{N}{2}\Lambda\widetilde{\Omega}_0(\widetilde{z},u_z).
 \label{eq2B.1}
\end{align}
\begin{figure*}[h]
        \centering
           \subfloat{%
              \includegraphics[scale=0.4]{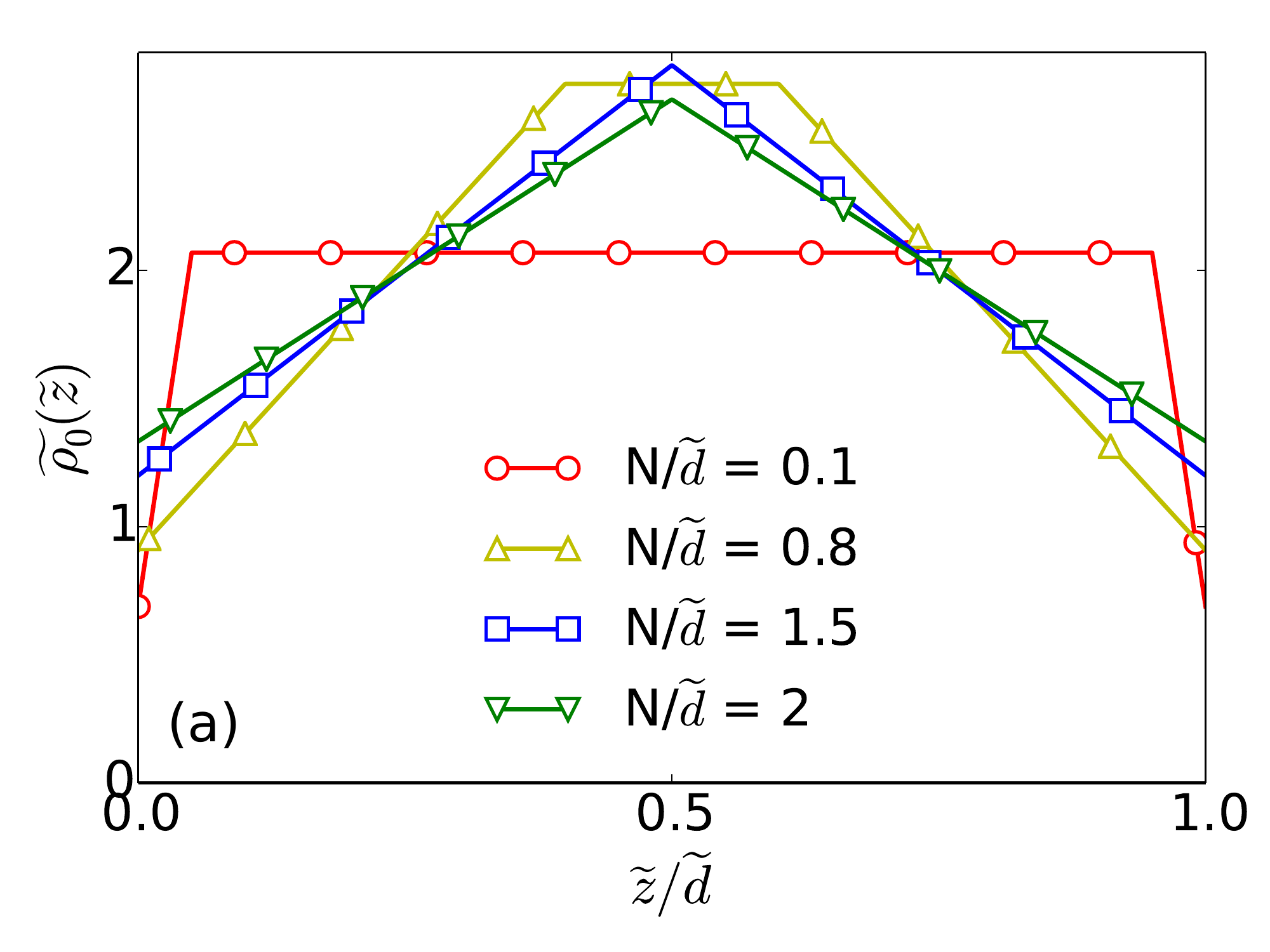}%
              \label{Fig4a}%
           } 
           \subfloat{%
              \includegraphics[scale=0.4]{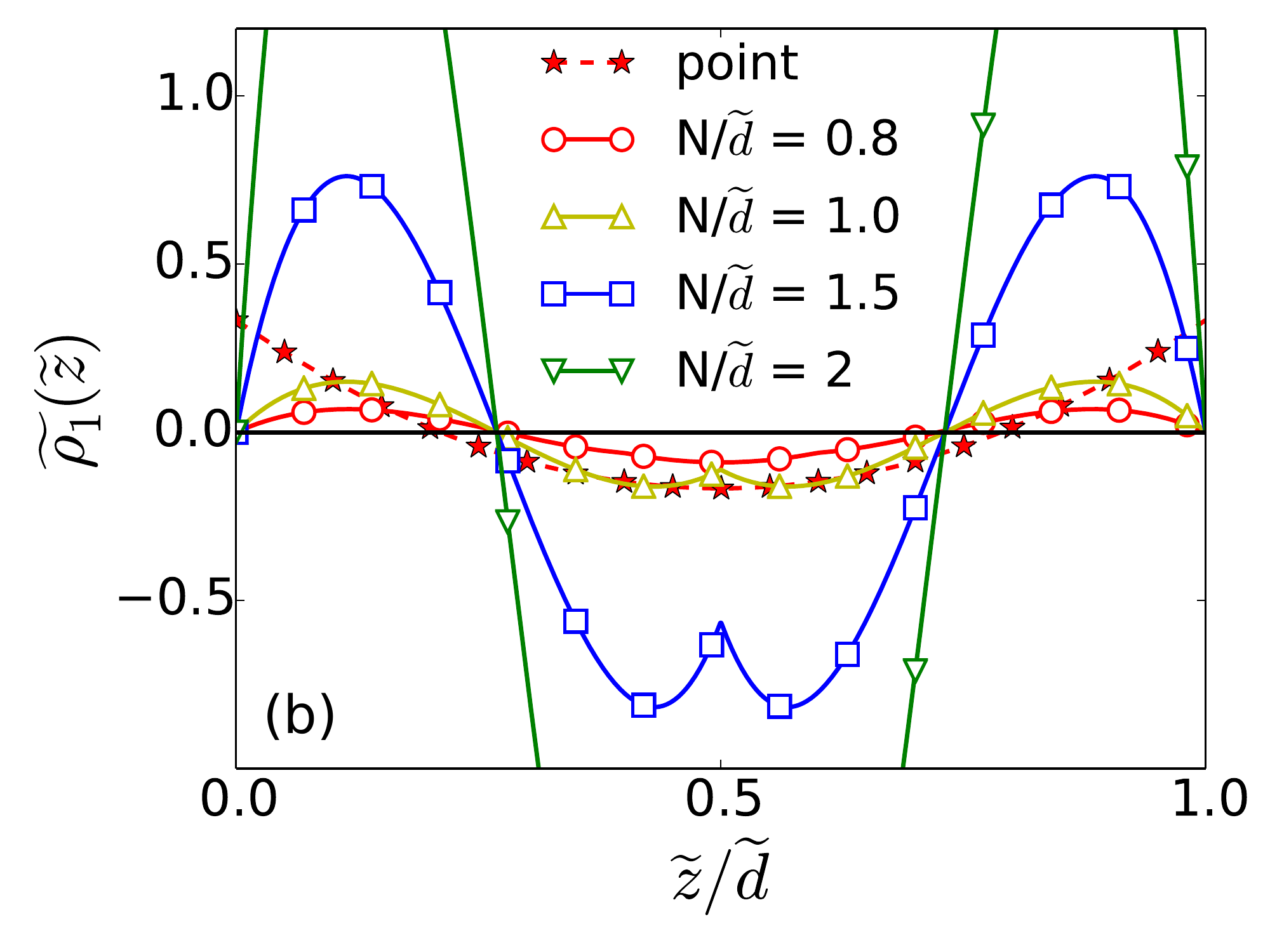}%
              \label{Fig4b}%
           }
           \caption{(Color online) Two wall: (a) Zeroth order density from equation \eqref{eq2B.2} for rodlike counterions confined between two walls separated 
           by a distance $\widetilde{d}$ is shown. 
           $\widetilde{z}$ is the position of the center of the rods. 
           (b) the corresponding leading order density term for different lengths of the rods and the corresponding point-particle case.}
           \label{Fig4}
 \end{figure*}
When $\widetilde{d} > N$, then using the explicit form of $\Omega_0(\widetilde{z},u_z)$ from equation \eqref{eq2A.3} we get the orientation averaged 
density, as defined in equation \eqref{eq2A.7}, to be
\begin{align}
 \widetilde{\rho}_0(\widetilde{z}) & =  \frac{1}{2}\Lambda_0 N\left(1+\frac{4\widetilde{z}}{N}\right)\hspace{15mm}\widetilde{z} < N/2, \nonumber \\
 & =  \frac{3}{2}\Lambda_0 N\hspace{34mm} N/2 < \widetilde{z} < \widetilde{d}-N/2, \nonumber \\
 & =  \frac{1}{2}\Lambda_0 N\left(1+\frac{4(\widetilde{d}-\widetilde{z})}{N}\right)\hspace{4mm} \widetilde{d}-N/2 < \widetilde{z} < \widetilde{d}. 
 \label{eq2B.2}
\end{align}
The normalization $\Lambda_0$ is obtained from the charge neutrality condition to be
\begin{equation}
 \Lambda_0 = \frac{4}{N(3\widetilde{d} - N)}.
 \label{eq2B.3}
\end{equation}
Similarly in case of $\widetilde{d} < N$, we get
\begin{align}
 \widetilde{\rho}_0(\widetilde{z}, u_z) & =  \frac{1}{2}\Lambda_0\left(1+\frac{4\widetilde{z}}{N}\right) \hspace{25mm}\widetilde{z} < \widetilde{d}/2, \nonumber \\
 & =  \frac{1}{2}\Lambda_0\left(1+\frac{4(\widetilde{d}-\widetilde{z})}{N}\right) \hspace{15mm}  \widetilde{d}/2 < \widetilde{z} < \widetilde{d}. 
 \label{eq2B.4}
\end{align}
 The charge neutrality condition yields $\Lambda_0 = 4N/\widetilde{d}(\widetilde{d}+N)$. Figure \ref{Fig4}-(a) plots the zeroth order orientation averaged density of the rods. 
 For very small rods, the density is nearly uniform similar to the point particle 
behavior. Since the interaction with the walls are zero ( $\widetilde{w}(\widetilde{z}) = 0$ ), the stronger repulsion
forces between the longer rodlike ions cause them to accumulate at the mid-plane of the two walls. The leading order density term has been calculated 
approximately in equation \eqref{eqE.6} in Appendix \ref{appendixE}. We plot it for different lengths of the rods in Figure \ref{Fig4}-(b). 
The leading order density term roughly measures the correlations between the rods. Figure \ref{Fig4}-(b) shows that the longer rods are more strongly correlated
than the shorter ones.

\begin{figure*}[h]
        \centering
           \subfloat{%
              \includegraphics[scale=0.4]{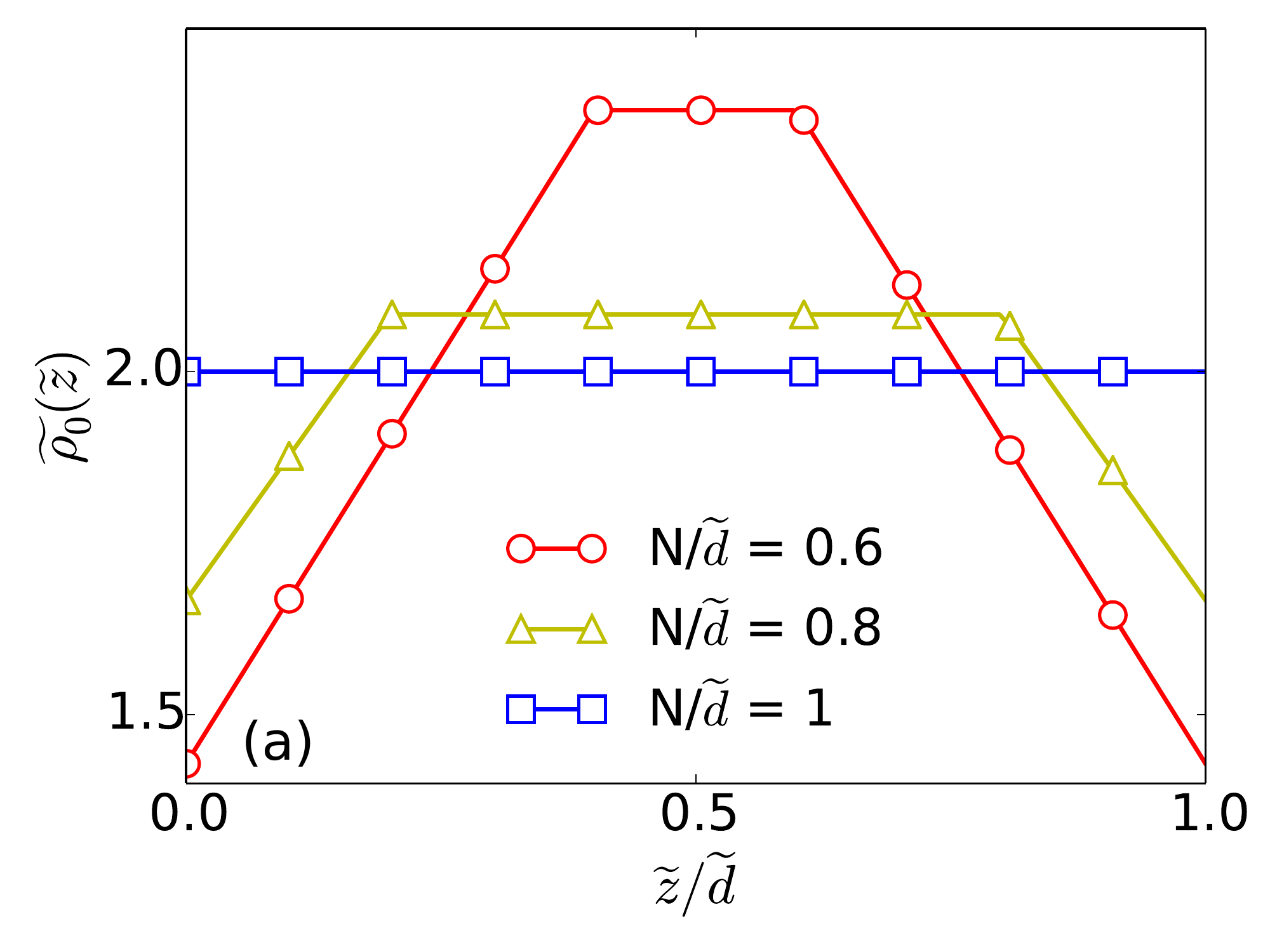}%
              \label{Fig5a}%
           } 
           \subfloat{%
              \includegraphics[scale=0.4]{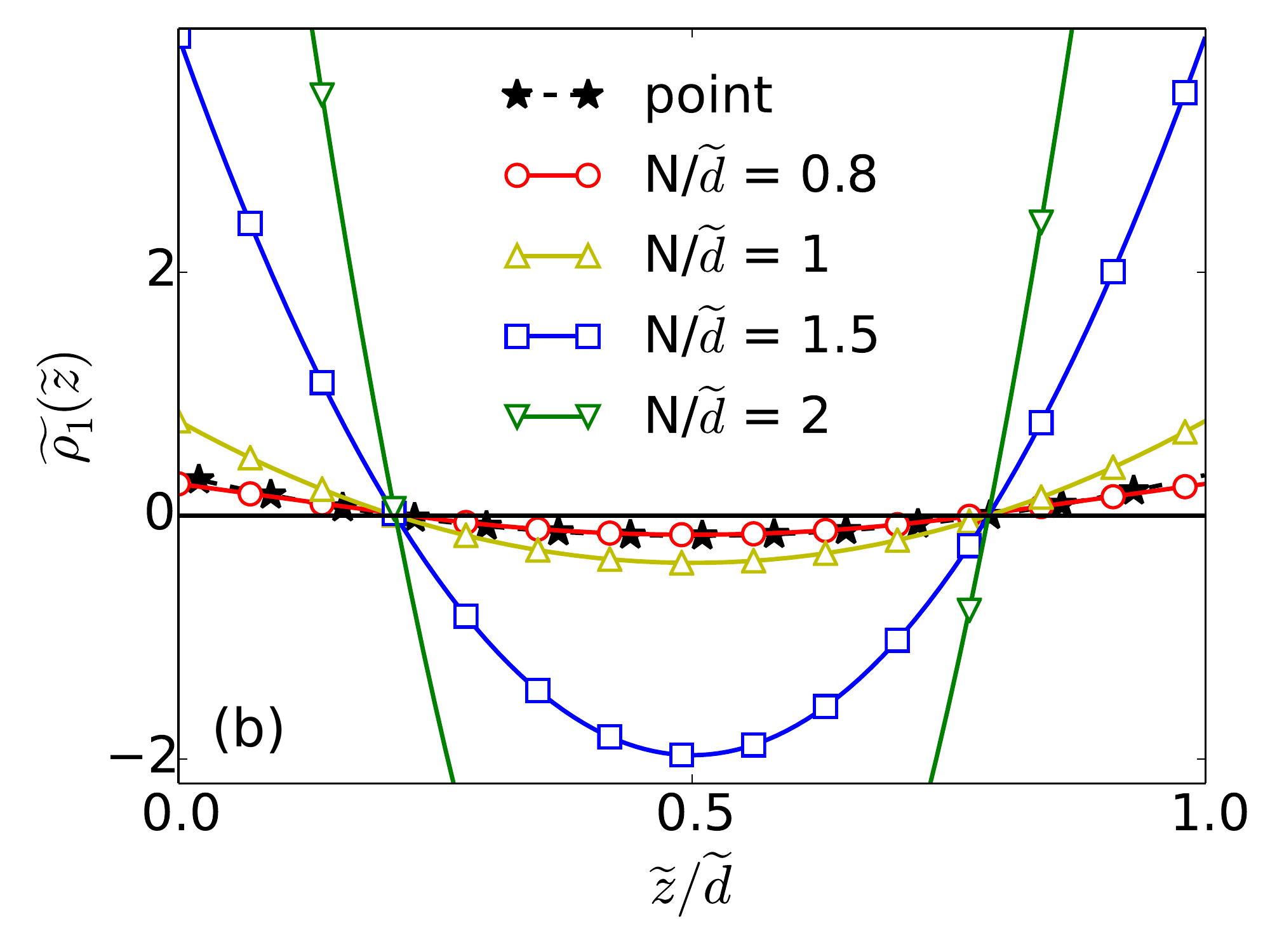}%
              \label{Fig5b}%
           }
           \caption{(Color online) Two wall: (a) Zeroth order density from equation \eqref{eq2B.5} for different lengths of the rods confined between
           two walls when $\widetilde{z}$ denotes the position 
           of one end of the rods.
           (b) The corresponding leading order density term for different lengths of the rods and the point-particle.}
           \label{Fig5}
 \end{figure*}
 
Now with the alternate parameterization with one end of the rods and using the excluded volume term defined in equation \eqref{eq2A.9},
we recover the results of Kim \textit{et al.} \cite{kim2008attractions} as shown in Figure \ref{Fig5}. When $N < \widetilde{d} < 2N$, then
\begin{align}
 \widetilde{\rho}_0(\widetilde{z}) & =  \frac{N}{2}\Lambda_0\left(1+\frac{\widetilde{z}}{N}\right)\hspace{15mm}\widetilde{z} < \widetilde{d} - N, \nonumber \\
 & =  \frac{1}{2}\widetilde{d}\Lambda_0\hspace{34mm} \widetilde{d} - N < \widetilde{z} < N, \nonumber \\
 & = \frac{N}{2}\Lambda_0\left(1+\frac{\widetilde{d} - \widetilde{z}}{N}\right)\hspace{15mm} N < \widetilde{z} < \widetilde{d}.
 \label{eq2B.5}
\end{align}
The normalization $\Lambda_0$ is obtained from the charge neutrality condition to be 
\begin{equation}
 \Lambda_0 = \frac{2}{N\left(\widetilde{d} - N/2\right)}.
 \label{eq2B.6}
\end{equation}
 
In case of $\widetilde{d} < N$, we have $\widetilde{\rho}_0(\widetilde{z}) = 2/\widetilde{d}$. Figures \ref{Fig5}-(a) and \ref{Fig5}-(b) shows the two lowest order density terms.
From Figure \ref{Fig5}-(a) we see that when the rods are longer than the separation between the walls, it loses the rotational degrees of freedom and looks more like point particles.
The leading order density in Figure \ref{Fig5}-(b) almost coincides with the point-particle when the length of the rod is $0.8$ times the distance between the walls. 

\begin{figure*}[h]
        \centering
              \includegraphics[scale=0.5,trim={0cm 0cm 0 0cm},clip]{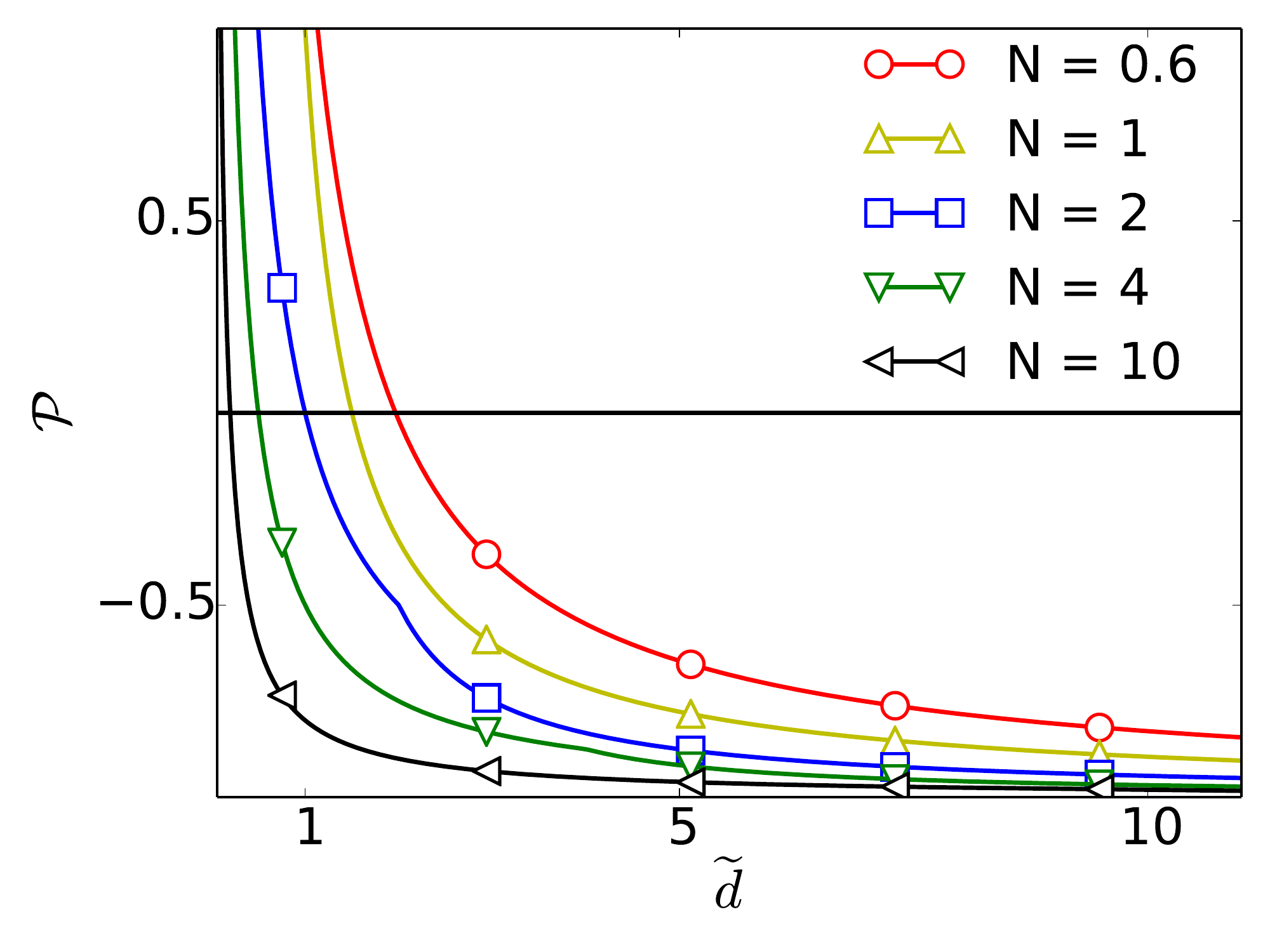}%           
           \caption{(Color online) Inter-wall pressure vs the inter-wall distance for various lengths of the rods.}
           \label{Fig6}
 \end{figure*}
 
 The pressure of the system is calculated from the contact value theorem which relates the pressure of the system to the interfacial density distribution of the 
 polymer counterions \cite{ kim2008attractions, israelachvili2011intermolecular}
\begin{equation}
\mathcal{P}  = \widetilde{\rho}_0(0) - 1. 
 \label{eq2B.7}
\end{equation}
 In the SC limit $\Xi \rightarrow \infty$ we obtain the pressure to be
\begin{align}
\mathcal{P} & = \frac{1}{N(\widetilde{d} - N/2)} - 1 \hspace{2mm} N < \widetilde{d} , \nonumber \\
 & = \frac{2}{N\widetilde{d}} - 1\hspace{5mm} \widetilde{d} < N.
 \label{eq2B.8}
\end{align}
Note that the factor of $1/N$ in the above calculations is due to the fact that only one end of the polymers are touching the surface. In Figure \ref{Fig6} we plot the 
pressure between the walls vs the inter-wall distance for different lengths of the rods. The longer rods have more charges and have stronger
electrostatic correlations compared to the shorter rods. They crowd near the walls and screen out the surface charge of the walls resulting in 
larger attractions between the walls than the shorter rods. In all the
experiments and simulations it have been found that the higher valency of the ions cause stronger like-charge attractions.

It would be instructive to compare our results to the earlier work on rodlike counterions confined between two walls by Bohinc \textit{et al.}
\cite{bohinc2012interactions}. Their densities resemble more closely to the densities we obtained for a single wall in Section \ref{subsectionA} which
is true when the rods are much smaller then the distance between the walls. However for the two wall case though we reproduce the results of Kim \textit{et al.}
\cite{kim2008attractions} our results disagrees with that of Bohinc \textit{at al.}. The pressure in Figure \ref{Fig6} shows a kink similar to that obtained
by Bohinc and his coworkers. In agreement with their result the kink occurs when the separation between the walls equals the length of rods 
(both $N = \widetilde{d} = 2$).

\section{conclusion and discussions}
\label{Sec3}
We have developed a polymer field theory for polyelectrolytic counterions in the presence of a fixed charge distribution in the SC regime. 
Whereas for the point charges the thermodynamics 
is described by a single parameter $\Xi$, the Coulomb coupling parameter, for polyelectrolytes we get an additional 
dependence on the length of the polymers that incorporates their geometry. Similarly the functional form of the thermodynamic quantities of polyelectrolytes
depends on the nature of the polymers themselves like the gaussian, wormlike or rodlike polymers. For rodlike polymers the analytical form of these 
functions/functionals can be computed exactly. 
Therefore we applied our field theoretic results to rodlike polyelectrolytes and consider two specific cases, when the rods are confined by one charged wall and
when they are confined between two charged walls. The walls restrict the orientational degrees of freedom of the rods close to them
which causes depletion near the wall and force them to align parallel to the surface of the walls. We parameterized the rodlike molecules in two different ways,
based on the center of the rods and based on one end of the rods, and have derived the thermodynamics in both cases. Because of their smaller size
the smaller polyelectrolytes can lower their free energy more easily by reorienting themselves near the wall than the larger ones. In the two wall case,
the densities are the same as obtained by Kim \textit{et al.} \cite{kim2008attractions} for dumbell-like counterions because of the same geometry
however our densities results disagree with Bohinc \textit{at al.} \cite{bohinc2012interactions}. However the inter-wall pressure in Bohinc \textit{at al.}
shows a kink when the inter-wall distance equals the rod length which is reproduced in our case.
Looking at the inter-wall pressure we find that when the walls are close they experience repulsion. However as their separation increases the interactions 
change to attractions. The longer polyelectrolytic rods have higher charges and higher correlations causing attractions at shorter inter-wall distances than 
smaller polyelectrolytes. The comparison with the other known limits, the point-particle systems \cite{Netz, moreira2001binding} and rodlike counterions
\cite{bohinc2012interactions} have also been explored. 

A recent formalism uses the fact the condensed counterions form a Wigner crystal like structure around the oppositely charged surface 
in the limit of large coupling parameter and shows the leading 
order correction to be of order $1/\sqrt{\Xi}$ not $1/\Xi$ as in case of Netz's formalism \cite{samajPRE,samajPRL}. 
This correction comes from the two particle interaction and the agreement with the Monte Carlo simulations is quite good. 
The zeorth order terms are the same as Netz formalism. The position dependent term of the leading order term of the density 
also remains same except the coefficient is different. For our model to work, the polymers should be shorter than
the size of the Wigner cell. In the case of rodlike polyelectrolytes, the two dimensional correlation 
structure on the macroion surface would be different from the Wigner lattice and the calculation of the two polymer energy is 
non-trivial because of the orientational degrees of freedom. In addition, the excluded volume effects in the leading order corrections
would be more important for polymers. The theory developed here can also be used to study the SC limit of other polymer models 
like Gaussian polyelectrolytes. This will be done in a subsequent paper.

\section{Acknowledgments}

YSJ is supported by the
Ministry of Education, Science, and Technology (NRF-
2012R1A1A2009275, NRF-C1ABA001-2011-0029960) of
the National Research 
Foundation of Korea (NRF).

\appendix

\section{Grand partition function in the SC limit}
\label{appendix0}

Starting from the Hamiltonian \eqref{eq1.7} we derive the expression \eqref{eq1.13} for the grand partition function. Since equation \eqref{eq1.9}
is Gaussian in density $\hat{\rho}$ we can applying the Hubbard-Stratonovich transformation \cite{hubbard1959calculation, stratonovich1957method} by 
introducing a field $\phi$
\begin{align}
 \mathcal{Z} & = \mathcal{Z}_0\sum\limits_{n=0}^{\infty}\frac{1}{n!}\left(\frac{\Lambda}{2\pi\Xi}\right)^n\int\left[\prod\limits_{i=1}^{n}\mathcal{D}\widetilde{\mathbf{r}}_i
  \widetilde{\Omega}(\widetilde{\mathbf{r}}_i)\right]\int\frac{\mathcal{D}\phi}{\mathcal{Z}_v}\exp\biggl(- \frac{1}{2\Xi}\int d\widetilde{\mathbf{x}}
  d\widetilde{\mathbf{x}}^{\prime}\phi(\widetilde{\mathbf{x}})\phi(\widetilde{\mathbf{x}}^{\prime})\widetilde{V}^{-1}(\vert\widetilde{\mathbf{x}}-
  \widetilde{\mathbf{x}}^{\prime}\vert) - \nonumber \\ &\int d\widetilde{\mathbf{x}}\widetilde{w}(\widetilde{\mathbf{x}})\hat{\rho}(\widetilde{\mathbf{x}}) +
   \int d\widetilde{\mathbf{x}}h(\widetilde{\mathbf{x}})\hat{\rho}(\widetilde{\mathbf{x}}) - i\int d\widetilde{\mathbf{x}}\phi(\widetilde{\mathbf{x}})
   \hat{\rho}(\widetilde{\mathbf{x}}) + \frac{\Xi}{2}nV_s -\sum_{i=1}^n\widetilde{H}_0^i \biggr),
 \label{eqA0.1}
\end{align}
where we have used the definition of $\mathcal{Z}_0$ in equation \eqref{eq1.12} and $\mathcal{Z}_v = (2\pi)^n\sqrt{\Det\widetilde{V}}$. We have used the scaled fugacity 
$\Lambda$ instead of the fugacity $\lambda$
in order to make the $1/\Xi$ dependence of the equation more explicit. The integration over the polymer fields $\{\widetilde{\mathbf{r}}_i\}$ can be 
performed to get
\begin{align}
 \mathcal{Z} & = \mathcal{Z}_0\sum\limits_{n=0}^{\infty}\frac{1}{n!}\left(\frac{\Lambda}{2\pi\Xi}\right)^n\int\frac{\mathcal{D}\phi}{\mathcal{Z}_v}\exp\biggl(- \frac{1}{2\Xi}\int d\widetilde{\mathbf{x}}
  d\widetilde{\mathbf{x}}^{\prime}\phi(\widetilde{\mathbf{x}})\phi(\widetilde{\mathbf{x}}^{\prime})\widetilde{V}^{-1}(\vert\widetilde{\mathbf{x}}-
  \widetilde{\mathbf{x}}^{\prime}\vert)\biggr)\times \nonumber\\& \int\left[\prod\limits_{i=1}^{n}\mathcal{D}\widetilde{\mathbf{r}}_i
  \widetilde{\Omega}(\widetilde{\mathbf{r}}_i)\right]\exp\biggl(-\int d\widetilde{\mathbf{x}}\hat{\rho}(\widetilde{\mathbf{x}})\left(\widetilde{w}(\widetilde{\mathbf{x}}) +
   i\phi(\widetilde{\mathbf{x}}) - h(\widetilde{\mathbf{x}})\right) + \frac{\Xi}{2}nV_s -\sum_{i=1}^n\widetilde{H}_0^i  \biggr) \nonumber \\
   &  = \mathcal{Z}_0\sum\limits_{n=0}^{\infty}\frac{1}{n!}\left(\frac{\Lambda}{2\pi\Xi}\right)^n\int\frac{\mathcal{D}\phi}{\mathcal{Z}_v}\exp\biggl(- \frac{1}{2\Xi}\int d\widetilde{\mathbf{x}}
  d\widetilde{\mathbf{x}}^{\prime}\phi(\widetilde{\mathbf{x}})\phi(\widetilde{\mathbf{x}}^{\prime})\widetilde{V}^{-1}(\vert\widetilde{\mathbf{x}}-
  \widetilde{\mathbf{x}}^{\prime}\vert)\biggr)Q_1^n 
 \label{eqA0.2}
\end{align}
where $Q_1$ is the single polymer parition function defined by
\begin{equation}
Q_1[h-\widetilde{w}-i\phi]  = \int\mathcal{D}\widetilde{\mathbf{r}}\widetilde{\Omega}(\widetilde{\mathbf{r}}) 
\exp\biggl(-\widetilde{H}_0[\widetilde{\mathbf{r}}]-\int_0^Nds\widetilde{w}(\widetilde{\mathbf{r}}(s)) -i\int_0^Nds\phi(\widetilde{\mathbf{r}}(s))
  +\int_0^Ndsh(\widetilde{\mathbf{r}}(s)) + \frac{\Xi}{2}V_s  \biggr).
 \label{eqA0.3}
\end{equation}
$\widetilde{H}_0[\widetilde{\mathbf{r}}]$ is the single polymer part of Hamiltonian. For Gaussian chains of average monomer length $a$ it would be
$3/(2a^2)\int_0^Nds\dot{\widetilde{\mathbf{r}}}(s)^2$. For wormlike chain of which the rodlike polymers with the bending rigidity $\kappa$ are a special case it 
is $\kappa\int_0^Nds\ddot{\widetilde{\mathbf{r}}}(s)^2$. Equation \eqref{eqA0.2} can be equivalently written in the form of equation \eqref{eq1.13}
\begin{align}
 \mathcal{Z} &  = \mathcal{Z}_0\sum_{j=0}^{\infty}\frac{1}{n!}\left(\frac{\Lambda}{2\pi\Xi}\right)^n\left\langle\widetilde{Q}_1^n[ h 
 -\widetilde{w}-i\phi] \right\rangle_{\phi} \nonumber\\
 & = \mathcal{Z}_0\left\langle\exp\left(\frac{\Lambda}{2\pi\Xi}\widetilde{Q}_1[h -\widetilde{w}-i\phi] \right)\right\rangle_{\phi} 
 \label{eqA0.4}
\end{align}
The averaging $\langle..\rangle_{\phi}$ is with respect to $ \frac{1}{2\Xi}\int_{\widetilde{\mathbf{x}},\widetilde{\mathbf{x}^{\prime}}}
\phi(\widetilde{\mathbf{x}})V^{-1}(\widetilde{\mathbf{x}} -\widetilde{\mathbf{x}^{\prime}})\phi(\widetilde{\mathbf{x}^{\prime}})$.

\section{SC leading order densities for rodlike counterions}
\label{appendixC}

Starting with equation \eqref{eq1.20} we derive an explicit formula for the leading order density term $\widetilde{\rho}_1$ using the 
form of the single polymer partition function $Q_1$ for rodlike polymers in equation \eqref{eq2.0}. We define two new quantities 
 \begin{align}
 \hat{\rho}_1(\widetilde{\mathbf{x}},\mathbf{u},\widetilde{\mathbf{y}}) & =\int_{0}^{N}ds\delta(\widetilde{\mathbf{x}}+s\mathbf{u}-
 \widetilde{\mathbf{y}}), \label{eqC.1}\\
 \hat{\rho}_2(\widetilde{\mathbf{x}},\mathbf{u},\widetilde{\mathbf{y}},s) & =\int_{0}^{N}ds^{\prime}\delta(\widetilde{\mathbf{x}}+
 (s^{\prime}-s)\mathbf{u}-\widetilde{\mathbf{y}}).
 \label{eqC.2}
 \end{align}
In terms of these $Q_1$ in equation \eqref{eq2.0} and the single rod density in equation \eqref{eq2.1} as
\begin{align}
 Q_1[\widetilde{w}] & = \frac{1}{4\pi}\int_{\Omega} \widetilde{\mathbf{x}}\int d\mathbf{u}\exp\left(-\int d\widetilde{\mathbf{y}}
 \hat{\rho}_1(\widetilde{\mathbf{x}},\mathbf{u},\widetilde{\mathbf{y}})\widetilde{w}(\widetilde{\mathbf{y}}) \right) \label{eqC.3} \\
 \widetilde{\rho}_0(\widetilde{\mathbf{x}},\mathbf{u}) & = \frac{\Lambda}{4\pi}\int_{0}^{N}ds\exp\left(-\int d\widetilde{\mathbf{y}}
 \hat{\rho}_2(\widetilde{\mathbf{x}},\mathbf{u},\widetilde{\mathbf{y}},s)\widetilde{w}(\widetilde{\mathbf{y}}) \right).
 \label{eqC.4}
 \end{align}
 
The leading order correction term to the density is obtained after performing the Gaussian
averaging in equation \eqref{eq1.21} using the single rod partition function in equation \eqref{eq2.0} and single rod density
in equation \eqref{eq2.1}
 \begin{align}
 \widetilde{\rho}_1(\widetilde{\mathbf{x}},\mathbf{u}) & = \frac{\Lambda^2}{32\pi^3}\int_{\Omega} d\widetilde{\mathbf{x}}^{\prime}
 \int d\mathbf{u}^{\prime}\int_{0}^{N} ds \biggl( \exp\biggl[ -\frac{1}{2}\int d\widetilde{\mathbf{y}}\left[
 \hat{\rho_1}(\widetilde{\mathbf{x}}^{\prime},\mathbf{u}^{\prime},\widetilde{\mathbf{y}})+\hat{\rho_2}(\widetilde{\mathbf{x}},\mathbf{u},\widetilde{\mathbf{y}},s)\right]\Xi\widetilde{V}(\widetilde{\mathbf{y}}-\widetilde{\mathbf{y}^{\prime}}) \nonumber\\
 \nonumber&\times\left[\hat{\rho_1}(\widetilde{\mathbf{x}}^{\prime},\mathbf{u}^{\prime},\widetilde{\mathbf{y}^{\prime}})+\hat{\rho_2}(\widetilde{\mathbf{x}},\mathbf{u},\widetilde{\mathbf{y}^{\prime}},s)\right]
  -\int d\widetilde{\mathbf{y}}\left(\hat{\rho_1}(\widetilde{\mathbf{x}}^{\prime},\mathbf{u}^{\prime},\widetilde{\mathbf{y}})+\hat{\rho_2}(\widetilde{\mathbf{x}},
  \mathbf{u},\widetilde{\mathbf{y}},s)\right)\widetilde{w}(\widetilde{\mathbf{y}}) + V_s\biggr] \nonumber\\
  & - \exp\biggl[ 
  -\int d\widetilde{\mathbf{y}}\left(\hat{\rho_1}(\widetilde{\mathbf{x}}^{\prime},\mathbf{u}^{\prime},\widetilde{\mathbf{y}})+\hat{\rho_2}(\widetilde{\mathbf{x}},
  \mathbf{u},\widetilde{\mathbf{y}},s)\right)\widetilde{w}(\widetilde{\mathbf{y}}) \biggr] \biggr)\nonumber\\
& = \frac{\Lambda^2}{32\pi^3}\int_{\Omega} d\widetilde{\mathbf{x}}^{\prime}\int d\mathbf{u}^{\prime}\int_{0}^{N} ds \exp\biggl[ 
  -\int d\widetilde{\mathbf{y}}\left(\hat{\rho_1}(\widetilde{\mathbf{x}}^{\prime},\mathbf{u}^{\prime},\widetilde{\mathbf{y}})+\hat{\rho_2}(\widetilde{\mathbf{x}},
  \mathbf{u},\widetilde{\mathbf{y}},s)\right)\widetilde{w}(\widetilde{\mathbf{y}}) \biggr] \times \nonumber \\ &\biggl[\exp\left(-\Xi\int ds^{\prime}ds^{\prime\prime}
  V\left(\widetilde{\mathbf{x}}+(s^{\prime}-s)\mathbf{u}-\widetilde{\mathbf{x}}^{\prime} - s^{\prime\prime}\mathbf{u}^{\prime}\right)\right)-1\biggr].
  \label{eqC.5}
  \end{align}
It is convenient to express the leading order density distributions in terms of the following quantities 
\begin{align}
q_1(\widetilde{\mathbf{x}},\mathbf{u}) & = \frac{1}{4\pi}\exp(-\int d\widetilde{\mathbf{y}}\hat{\rho_1}(\widetilde{\mathbf{x}},\mathbf{u},\widetilde{\mathbf{y}})
\widetilde{w}(\widetilde{\mathbf{y}})), \label{eqC.6}\\
q_2(\widetilde{\mathbf{x}},\mathbf{u},s) & = \frac{1}{4\pi}\exp(-\int d\widetilde{\mathbf{y}}\hat{\rho_2}(\widetilde{\mathbf{x}},\mathbf{u},s,\widetilde{\mathbf{y}})
\widetilde{w}(\widetilde{\mathbf{y}})).
\label{eqC.7}
\end{align}
We can then write $\widetilde{\rho}_0$ and $\widetilde{\rho}_1$ as
\begin{align}
  \widetilde{\rho_0}(\widetilde{\mathbf{x}},\mathbf{u}) & = \Lambda \int_{0}^{N} ds q_2(\widetilde{\mathbf{x}},\mathbf{u},s), \label{eqC.8}\\
 \widetilde{\rho_1}(\widetilde{\mathbf{x}},\mathbf{u}) & = \frac{\Lambda^2}{2\pi}\int_{\Omega} d\widetilde{\mathbf{x}}^{\prime}\int d\mathbf{u}^{\prime}\int_{0}^{N} ds 
 q_1(\widetilde{\mathbf{x}}^{\prime},\mathbf{u}^{\prime})q_2(\widetilde{\mathbf{x}},\mathbf{u},s)\biggl[\exp\biggl(-\Xi\int ds^{\prime}ds^{\prime\prime}
  V\biggl(\widetilde{\mathbf{x}}+(s^{\prime}-s)\mathbf{u}-\widetilde{\mathbf{x}}^{\prime} \nonumber\\& - s^{\prime\prime}\mathbf{u}^{\prime}\biggr)\biggr)-1\biggr]. 
  \label{eqC.9}
  \end{align}
  
\section{Leading order fugacities $\Lambda_0$ and $\Lambda_1$ for rodlike molecules}
\label{appendixA}

The terms in the expansion of the fugacity in powers of $1/\Xi$ in equation \eqref{eq2A.3.1} is determined from the normalization
of the density
\begin{equation}
 \int_{\Omega} d\widetilde{\mathbf{x}}\widetilde{\rho}(\widetilde{\mathbf{x}}) = n_w,
 \label{eqA.1}
\end{equation}
where $n_w$ is the number of walls in the confinement. Expanding $\Lambda$ and $\widetilde{\rho}$ in powers of $1/\Xi$ the 
above equation becomes
\begin{equation}
 \int_{\Omega} d\widetilde{\mathbf{x}}\Lambda_0\widetilde{\rho}_0(\widetilde{\mathbf{x}}) + \frac{1}{\Xi}\int_{\Omega} d\widetilde{\mathbf{x}}\left(\Lambda_1\widetilde{\rho}_0(\widetilde{\mathbf{x}})
 +\frac{\Lambda_0^2}{\pi}\widetilde{\rho}_1(\widetilde{\mathbf{x}})\right) = n_w.
 \label{eqA.2}
\end{equation}
We solve this equation for $\Lambda_0$ and $\Lambda_1$ for some given forms for $\widetilde{\rho}_0$ and $\widetilde{\rho}_1$
\begin{align}
 \Lambda_0 & = \frac{n_w}{\int_{\Omega} d\widetilde{\mathbf{x}}\widetilde{\rho}_0(\widetilde{\mathbf{x}})}, \label{eqA.3}\\
 \Lambda_1 & = -\Lambda_0^2\frac{\int_{\Omega} d\widetilde{\mathbf{x}}\widetilde{\rho}_1(\widetilde{\mathbf{x}})}{\int_{\Omega} d\widetilde{\mathbf{x}}\widetilde{\rho}_0(\widetilde{\mathbf{x}})}
 = -\frac{\Lambda_0^3}{n_w}\int_{\Omega} d\widetilde{\mathbf{x}}\widetilde{\rho}_1(\widetilde{\mathbf{x}})
 =-n_w^2\frac{\int_{\Omega} d\widetilde{\mathbf{x}}\widetilde{\rho}_1(\widetilde{\mathbf{x}})}{\left(\int_{\Omega} d\widetilde{\mathbf{x}}
 \widetilde{\rho}_0(\widetilde{\mathbf{x}})\right)^3}.
\label{eqA.4}
 \end{align}

In particular for rodlike polymers, using the expressions for $\widetilde{\rho}_0$ and $\widetilde{\rho}_1$ in equations \eqref{eq2.2} and \eqref{eq2.3}, we get
\begin{align}
 \Lambda_0 & = n_w\biggl/\left(\int_{\Omega} d\widetilde{\mathbf{x}}\int d\mathbf{u} \int_{0}^{N} ds 
 q_2(\widetilde{\mathbf{x}},\mathbf{u}, s)\right), \label{eqA.5}\\
 \Lambda_1 & = -\frac{\Lambda_0^2}{\iiint q_2(\widetilde{\mathbf{x}},\mathbf{u}, s)}\iiint q_1(\widetilde{\mathbf{x}^{\prime}},\mathbf{u}^{\prime})q_2(\widetilde{\mathbf{x}},\mathbf{u},s)\biggl[\exp\biggl(-\Xi\int ds^{\prime}ds^{\prime\prime}
  \widetilde{V}\biggl(\widetilde{\mathbf{x}}+(s^{\prime}-s)\mathbf{u}-\widetilde{\mathbf{x}}^{\prime} \nonumber\\& - s^{\prime\prime}\mathbf{u}^{\prime}\biggr)\biggr)-1\biggr],
  \label{eqA.6}
\end{align}
where we have used the notation $\iiint$ for integrals over all the variables $\widetilde{\mathbf{x}}$, $\mathbf{u}$ and $s$ and the excluded volume term 
$\widetilde{\Omega}(\widetilde{\mathbf{x}},\mathbf{u})$ has been included in the integral notation.

\section{Free energy}
\label{appendixB}

The free energy is defined in terms of the grand partition function \cite{naji2005counterions}
\begin{equation}
 \beta\mathcal{F}_N = N\ln\Lambda -\ln\mathcal{Z} ,
 \label{eqB.1}
\end{equation}
where $N = \Lambda\frac{\partial\ln\mathcal{Z}}{\partial\Lambda}$. In the large coupling limit, using the expansion of the fugacity as in equation \eqref{eq2A.3.1}
\begin{equation}
 \Lambda = \Lambda_0 + \frac{\Lambda_1}{\Xi} + \frac{\Lambda_2}{\Xi^2} + \mathcal{O}(1/\Xi^3)
 \label{eqB.2}
\end{equation}
and the expansion of the logarithm of the grand partition function in equation \eqref{eq1.17}
\begin{align}
 \ln\mathcal{Z} & = \ln\mathcal{Z}_0 + \left(\sum_{j=1}\frac{1}{j!}\left(\frac{\Lambda}{2\pi\Xi}\right)^jQ_j\right) - \frac{1}{2}\left(\sum_{j=1}\frac{1}{j!}
 \left(\frac{\Lambda}{2\pi\Xi}\right)^jQ_j\right)^2 \nonumber\\
 & = \ln\mathcal{Z}_0 + \left(\frac{\Lambda}{2\pi\Xi}\right)Q_1 + \frac{1}{2}\left(\frac{\Lambda}{2\pi\Xi}\right)^2(Q_2 - Q_1^2) + .. ,
 \label{eqB.3}
\end{align}
where $\ln\mathcal{Z}_0 = - \frac{1}{8\pi^2\Xi}\int d\widetilde{\mathbf{x}}d\widetilde{\mathbf{x}}^{\prime}\widetilde{\sigma}(\widetilde{\mathbf{x}})
\widetilde{V}(\widetilde{\mathbf{x}}-\widetilde{\mathbf{x}}^{\prime})\widetilde{\sigma}(\widetilde{\mathbf{x}}^{\prime})$. Taking the derivative 
of the above equation with respect to $\Lambda$ we get 
\begin{align}
 N & =  \left(\frac{\Lambda}{2\pi\Xi}\right)Q_1 + \left(\frac{\Lambda}{2\pi\Xi}\right)^2(Q_2 - Q_1^2) + .. \nonumber\\
 & = \frac{\Lambda_0}{2\pi\Xi}Q_1 + \frac{1}{2\pi\Xi^2}\left(\Lambda_1Q_1 + \frac{\Lambda_0^2}{2\pi}(Q_2-Q_1^2)\right)
 \label{eqB.4}
\end{align}

We expand the free energy in the inverse powers of $\Xi$
\begin{align}
 \mathcal{F} & = \frac{\mathcal{F}_1}{\Xi} + \frac{\mathcal{F}_2}{\Xi^2} + \mathcal{O}(1/\Xi^3).
 \label{eqB.5}
\end{align}
Using equations \eqref{eqB.3} and \eqref{eqB.4} in equation \eqref{eqB.1} we can readily solve the leading order term 
\begin{align}
 \mathcal{F}_1 & = \frac{\Lambda_0}{2\pi}Q_1\left(\ln\Lambda_0 -1\right) - \Xi\ln\mathcal{Z}_0, 
\label{eqB.6}
 \end{align}
where $-\ln\mathcal{Z}_0$ is the free energy of the fixed charges and for rodlike molecules $Q_1$ read
\begin{align}
 Q_1  & = \int_{\Omega} d\widetilde{\mathbf{x}}d\mathbf{u} q_1(\widetilde{\mathbf{x}},\mathbf{u}). 
  \label{eqB.7}
\end{align}
Thus the free energy of the counterions only is 
\begin{equation}
  \triangle\mathcal{F}_1  = \frac{\Lambda_0}{2\pi}Q_1\left(\ln\Lambda_0 -1\right) .
\label{eqB.8}
  \end{equation}

\section{Rodlike counterions confined by a charged wall}
\label{appendixD}
For rodlike molecules confined on the upper $z$-half plane by a charged wall, using the rescaled potential in equation \eqref{eq2A.1} in equation \eqref{eq2.4}
and \eqref{eq2.5} we get
\begin{align}
q_1(\widetilde{z}) & = \frac{1}{2}\exp(-N\widetilde{z}), \label{eqD.1}\\
q_2(\widetilde{z},u_z, s) &  = \frac{1}{2}\exp\left[-N(\widetilde{z}-su_z)\right] 
\label{eqD.2}
\end{align}
 in equation \eqref{eqA.5} we get the lowest order fugacity term as
\begin{align}
 \Lambda_0  = 1\biggr/\biggl(\frac{1}{2}+\frac{1}{N^2}\biggl(\gamma - \expIntEi(-N^2) + \ln N^2\biggr)\biggr).
\label{dqD.3}
 \end{align}

The first order term in the density function $\widetilde{\rho}_1$ averaged over the orientations in equation \eqref{eq2.3} becomes
\begin{align}
 \widetilde{\rho_1}(\widetilde{z}) 
   & = \frac{\Lambda^2}{\pi}\int_0^{\infty} d\widetilde{z}^{\prime} \int_{-1}^{1}du_z^{\prime}\widetilde{\Omega}_0(\widetilde{z}^{\prime},u_z^{\prime})
   \int_{-N/2}^{N/2} ds \int_{-1}^{1}du_z
   \widetilde{\Omega}_0(\widetilde{z},u_z)q_1(\widetilde{z}^{\prime})q_2(\widetilde{z},u_z,s)\int_0^{\infty}\rho d\rho \times\nonumber\\&
   \biggl[\exp\biggl(-\int ds^{\prime}ds^{\prime\prime}
  \frac{\Xi}{\sqrt{\rho^2 + \vert\widetilde{z}^{\prime}-\widetilde{z}_s\vert^2}}\biggr)-1\biggr] , 
  \label{eqD.4}
  \end{align}
where $\widetilde{z}_s(\widetilde{z},u_z,u^{\prime}_z,s^{\prime},s^{\prime\prime}) =  \widetilde{z} + (s^{\prime}-s)u_z - s^{\prime\prime}u^{\prime}_z $. The $s^{\prime}$ and
$s^{\prime\prime}$ integrals inside the exponential are a result of complicated many body interactions because of the elongated geometry of the rods and makes
the interactions non-local. The excluded volume factors $\widetilde{\Omega}_0$ further complicates the calculation of the above expression. To get some understanding
about the first order term we introduce some approximations to make the algebra tractable. For the first approximation we bring the $s^{\prime}$ and $s^{\prime\prime}$
outside the exponential function to make the interactions local. Next we remove the constraint imposed by the excluded volume terms and put them back at the end of the calculations. 
After making a change of variables we have
  \begin{align}
   \widetilde{\rho_1}(\widetilde{z})  & \approx \Lambda^2\int_0^{\infty} d\widetilde{z}^{\prime} \int_{-1}^{1}du_z^{\prime}\int_{-N/2}^{N/2} ds \int_{-1}^{1}du_z
   q_1(\widetilde{z}^{\prime})q_2(\widetilde{z},u_z,s)\int ds^{\prime}ds^{\prime\prime}
 \int_{\vert\widetilde{z}^{\prime}-\widetilde{z}_s\vert}^{\infty}x dx \times\nonumber\\&\biggl[e^{-\Xi/x}-1\biggr] \nonumber\\
 & = \frac{1}{N}\Lambda^2\iiint\limits_{s,u_z}q_2(\widetilde{z},u_z,s)\biggl[q_1(\widetilde{z}_s-t)
 \int_{t}^{\infty}dxx\left[e^{-\Xi/t}-1\right]\biggr\vert_0^{\widetilde{z}_s} +  \int_0^{\widetilde{z}_s}dt q_1(\widetilde{z}_s-t)
 t[e^{-\Xi/t}\nonumber\\ &-1]  - q_1(\widetilde{z}_s+t)\int_{t}^{\infty}dxx\left[\exp(-\Xi/x)-1\right]\biggr\vert_0^{\infty} -
 \int_0^{\infty}dt q_1(\widetilde{z}_s+t)t\left[e^{-\Xi/t}-1\right]\biggr] \nonumber\\ 
 & = \frac{1}{N}\Lambda^2\iiint\limits_{s,u_z}q_2(\widetilde{z},u_z,s)\biggl[q_1(0)
 \int_{\widetilde{z}_s}^{\infty}dxx\left[e^{-\Xi/t}-1\right] +  \int_0^{\widetilde{z}_s}dt q_1(\widetilde{z}_s-t) 
  t\left[e^{-\Xi/t}-1\right] - \nonumber\\ & \int_0^{\infty}dtq_1(\widetilde{z}_s+t)t\left[e^{-\Xi/t}-1\right]\biggr] \nonumber\\
 & = \frac{1}{N}\Lambda^2\iiint\limits_{s,u_z} q_2(\widetilde{z},u_z,s)\biggl[\int_0^{\infty}dt \left[q_1(\widetilde{z}_s-t)-q_1(\widetilde{z}_s+t)\right]
 t\biggl[e^{-\Xi/t} -1\biggr]  + \int_{\widetilde{z}_s}^{\infty}dt[q_1(0) \nonumber\\& - q_1(\widetilde{z}_s-t)]
  t\left[e^{-\Xi/t}-1\right]\biggr].
  \label{eqD.5}
 \end{align}
 The first order term of the fugacity is 
  \begin{align}
   \Lambda_1 & = -\frac{\Lambda_0^2}{N\iiint q_2(\widetilde{z}^{\prime},u_z^{\prime},s^{\prime})}\int dsd\widetilde{z}^{\prime}du_z^{\prime}
  q_2(\widetilde{z}^{\prime},u_z^{\prime},s) \int ds^{\prime}ds^{\prime\prime}\int du^{\prime\prime}_z\biggl[q_1(0)
 \int_{\widetilde{z}_s}^{\infty}dxx\times \nonumber\\ & [e^{-\Xi/x} -1]   +  \int_0^{\widetilde{z}_s}dt q_1(\widetilde{z}_s-t) 
  t\left[e^{-\Xi/x}-1\right] - \int_0^{\infty}dtq_1(\widetilde{z}_s+t)t\left[e^{-\Xi/x}-1\right]\biggr].
  \label{eqD.6}
  \end{align}

In the limit of $\Xi\rightarrow\infty$ after the expansion of the fugacity in powers of $1/\Xi$ and keeping only it's zeroth order term
\begin{align}
  \widetilde{\rho_1}(\widetilde{z}) & = \frac{1}{N}\Lambda_0^2\int dsdu_zq_2(\widetilde{z},u_z,s)\int ds^{\prime}ds^{\prime\prime}
 \int du^{\prime}\biggl[q_1(0)\int_{\widetilde{z}_s}^{\infty}dt t + q_1(\widetilde{z}_s)\int_0^{\widetilde{z}_s}dtt\exp(Nt)
  \nonumber\\&   - q_1(\widetilde{z}_s)\int_{0}^{\infty}dtt\exp(-Nt) \biggr] \nonumber\\
  & = \frac{1}{N}\Lambda_0^2\int du_zdsq_2(\widetilde{z},u_z,s)\int ds^{\prime}ds^{\prime\prime}
 \int du^{\prime}_zq_1(0)\biggl[\left(I-\frac{1}{2}\widetilde{z}_s^2\right) + \frac{1}{N^2}(\widetilde{z}_sN-1)\biggr].
\label{eqD.7}
 \end{align}
This expression is divergent as we claimed in Section \ref{Sec1} because $I \rightarrow \infty$. Similarly for the first order fugacity term in equation \eqref{eqD.6}
we have in the limit $\Xi\rightarrow\infty$
\begin{align}
  \Lambda_1 & = -\frac{\Lambda_0^2}{N\iiint q_2(\widetilde{z}^{\prime},u_z^{\prime},s^{\prime})}\int dsd\widetilde{z}^{\prime}
  du_z^{\prime}q_2(\widetilde{z}^{\prime},u_z^{\prime},s)\int ds^{\prime}ds^{\prime\prime}
 \int du_z^{\prime\prime}q_1(0)\biggl[(I-\frac{1}{2}\widetilde{z}_s^2) \nonumber\\ & + \frac{1}{N^2}(\widetilde{z}_sN-1)  
 \biggr] .
 \label{eqD.8}
\end{align}

The renormalized leading order term of the density becomes
\begin{align}
\widetilde{\rho_1}^{Ren}(\widetilde{z}) & = \Lambda_1 \widetilde{\rho}_0(\widetilde{z}) + \frac{\Lambda_0^2}{2\pi}
\widetilde{\rho}_1(\widetilde{z}).
\label{eqD.9}
  \end{align}
  In the limit $\Xi\rightarrow\infty$ the renormalized density becomes
  \begin{align}
 {\widetilde{\rho_1}}^{Ren}(\widetilde{z}) & = \frac{1}{N}\Lambda_0^2\int dsdu_z\widetilde{\Omega}_0(\widetilde{z},u_z)q_2(\widetilde{z},u_z,s)\int ds^{\prime}ds^{\prime\prime}
 \int du^{\prime}_zq_1(0)\biggl[\frac{1}{2}\left({\widetilde{z}_s}^2 - \langle\widetilde{z}_s^2\rangle\right) - 
 \frac{1}{N}\left(\widetilde{z}_s - \langle\widetilde{z}_s\rangle\right)\biggr] ,
\label{eqD.10}
 \end{align}
where we have put the excluded volume term $\widetilde{\Omega}_0(\widetilde{z},u_z)$ back into the density expression. The averaging $\langle ..\rangle$ is with respect 
to the weight $\int d\widetilde{z}^{\prime} du_z^{\prime}ds^{\prime}\widetilde{\Omega}_0(\widetilde{z}^{\prime},u_z^{\prime})q_2(\widetilde{z}^{\prime},u_z^{\prime},
s^{\prime})$ and $\widetilde{z}_s =  \widetilde{z} + (s^{\prime}-s)u_z -s^{\prime\prime}u^{\prime}_z$. Note that the final expression of the first order density has the same form 
as point particles except for the integrals over the rod contours $s$.

\section{Rodlike counterions confined between two charged walls}
\label{appendixE}
For rodlike molecules confined between two charged walls separated by a distance $\widetilde{d}$ we get from equations \eqref{eq2.4} and \eqref{eq2.5}
\begin{align}
q_1(\widetilde{z},u_z) & = \frac{1}{2} = q_2(\widetilde{z},u_z, s) .
\label{eqE.1}
\end{align}
We have used the fact that the electric field or energy between the walls are zero.
The first order term of the density averaged over the orientations is obtained following the same procedure as outlined in Appendix \ref{appendixD}
\begin{align}
\widetilde{\rho_1}(\widetilde{z}) 
   & = \frac{\Lambda^2}{4}\int_{-1}^1 du_z\widetilde{\Omega}_0(\widetilde{z},u_z) \int_{-1}^1 du_z^{\prime}\int_{-N/2}^{N/2} ds 
 \int_0^{\infty}\rho d\rho \biggl[\exp\biggl(-\int ds^{\prime}ds^{\prime\prime}\nonumber\\&
  \frac{\Xi}{\sqrt{\rho^2 + \vert\widetilde{z}-\widetilde{z}_s\vert^2}}\biggr)-1\biggr] .
  \label{eqE.2}
  \end{align}
Again we neglect the non-local the many-body interactions and keep only the local interactions. The $s^{\prime}$
and $s^{\prime\prime}$ integrals then can be taken out of the exponential and we get for the density expression \cite{Netz, moreira2001binding}
\begin{align}
 \widetilde{\rho_1}(\widetilde{z}) 
   & = \frac{\Lambda^2_0}{4}\int_{-1}^1 du_z\widetilde{\Omega}_0(\widetilde{z},u_z) \int ds ds^{\prime}ds^{\prime}\int du_z^{\prime}\int\biggl[\widetilde{d} \int_0^{\infty}dxx[e^{-\Xi/x}-1]
   + \int_0^{\widetilde{z}_s}dx(x-\widetilde{z}_s)x[e^{-\Xi/x} \nonumber \\& -1] + \int_0^{\widetilde{d}-\widetilde{z}_s}dx(x-\widetilde{d}+\widetilde{z}_s)x[e^{-\Xi/x}-1]\biggr]
\label{eqE.3}
   \end{align}
Similarly the first order term in fugacity is given by
\begin{align}
\Lambda_1
   & = -\frac{\Lambda^2_0}{4\int d\widetilde{z}du_zds\widetilde{\Omega}_0(\widetilde{z},u_z)q_2(\widetilde{z},u_z)}\int_0^{\widetilde{d}} d\widetilde{z}^{\prime}\int_{-1}^1 du_z^{\prime}\widetilde{\Omega}_0(\widetilde{z}^{\prime},u_z^{\prime})\int ds^{\prime}ds^{\prime\prime}\int du_z^{\prime\prime}
   \biggl[\widetilde{d} \int_0^{\infty}dxx[e^{-\Xi/x} \nonumber \\& -1] + 
   \int_0^{\widetilde{z}_s^{\prime}}dx(x-\widetilde{z}_s^{\prime}) x[e^{-\Xi/x}  -1] + \int_0^{\widetilde{d}-\widetilde{z}_s^{\prime}}dx(x-\widetilde{d}+
   \widetilde{z}_s^{\prime})x[e^{-\Xi/x}-1]\biggr].
 \label{eqE.4}
 \end{align}

In the limit of $\Xi\rightarrow\infty$ the first order term of density in equation \eqref{eqE.3} and fugacity \eqref{eqE.4} becomes
\begin{align}
 &\widetilde{\rho_1}(\widetilde{z}) 
    = -\frac{\Lambda^2_0}{4}\int_{-1}^1 du_zdu_z^{\prime}\widetilde{\Omega}_0(\widetilde{z},u_z)\int dsds^{\prime}ds^{\prime\prime}\biggl[\widetilde{d}I + 
   \frac{1}{6}\left((\widetilde{z}_s-\widetilde{d})^3-\widetilde{z}_s^3\right)\biggr], \nonumber\\
  & \Lambda_1  = \frac{\Lambda^2_0}{2\int d\widetilde{z}du_zds\widetilde{\Omega}_0(\widetilde{z},u_z)}\int_0^{\widetilde{d}} d\widetilde{z}^{\prime}\int_{-1}^1 du_z^{\prime}du_z^{\prime\prime}
   \widetilde{\Omega}_0(\widetilde{z}^{\prime},u_z^{\prime})
   \int dsds^{\prime}ds^{\prime\prime} \biggl[\widetilde{d}I + \frac{1}{6}\left((\widetilde{z}_s^{\prime}-\widetilde{d})^3-(\widetilde{z}_s^{\prime})^3\right)\biggr].
\label{eqE.5}
   \end{align}
Both the terms diverges, as $I \rightarrow \infty$. However the renormalized first order density defined in equation \eqref{eqD.9} remains finite because of 
the cancellation of the terms containing $I$
\begin{align}
 \widetilde{\rho_1}(\widetilde{z}) 
   & = -\frac{\Lambda_0^2}{4}\iiint\limits_{s,u_z}\biggl[\frac{1}{6}\left((\widetilde{z}_s-\widetilde{d})^3-\widetilde{z}_s^3\right)\biggr] +
   \frac{\Lambda_0^2}{4\widetilde{d}}\int_0^{\widetilde{d}}d\widetilde{z}^{\prime}\iiint\limits_{s,u_z}\biggl[\frac{1}{6}\left((\widetilde{z}_s^{\prime}-
   \widetilde{d})^3-(\widetilde{z}_s^{\prime})^3\right)\biggr] ,
\label{eqE.6}
   \end{align}
where $\widetilde{z}_s =  \widetilde{z} + (s^{\prime}-s)u_z - s^{\prime\prime}u^{\prime}_z$. The triple integral $\iiint$ denotes the integrals
over $s$ and $u_z$. Without the excluded volume effects with the walls $\widetilde{\Omega}_0 = 1$, we recover the single particle results of Netz \cite{Netz} 
for $N = 1$
\begin{equation}
 \widetilde{\rho_1}(\widetilde{z}) = \frac{2N^3}{\widetilde{d}}\left[\left(\widetilde{z}-\widetilde{d}/2\right)^2 - \frac{1}{12}\widetilde{d}^2 \right].
\label{eqE.7}
 \end{equation}

\bibliography{bibliography}

\end{document}